\documentclass[11pt]{article}
\usepackage[margin=1in]{geometry}
\usepackage{amsmath,amssymb,amsthm,mathtools}
\usepackage{bbm}
\usepackage{booktabs}
\usepackage{microtype}
\usepackage[colorlinks=true,linkcolor=blue,citecolor=blue,urlcolor=blue]{hyperref}
\usepackage{comment}
\usepackage{xcolor}
\usepackage{bm}
\usepackage{array}
\usepackage{float}
\usepackage[T1]{fontenc}
\usepackage{lmodern}
\usepackage{authblk}
\usepackage{standalone}
\usepackage{amsmath,amssymb,pgfplots,caption}
\pgfplotsset{compat=1.18}
\usepgfplotslibrary{groupplots}
\usepackage{threeparttable}
\usepackage{comment}

\newtheorem{theorem}{Theorem}[section]
\theoremstyle{plain}
\theoremstyle{plain}
\newtheorem{lemma}{Lemma}[subsection]
\newtheorem{proposition}{Proposition}[subsection]
\theoremstyle{plain}
\newtheorem{corollary}{Corollary}[subsection]

\theoremstyle{definition}\theoremstyle{plain}
\newtheorem{definition}{Definition}[subsection]
\newtheorem{remark}{Remark}[subsection]

\newcommand{\Tr}{\operatorname{Tr}}
\newcommand{\OPT}{\operatorname{OPT}}

\newcommand{\E}{\mathbb{E}}

\newcommand{\R}{\mathbb{R}}

\newcommand{\Hf}{H_{\mathrm{FMC}}}

\newcommand{\rgw}{r_{\mathrm{GW}}}

\definecolor{darkorange}{rgb}{1.0, 0.4, 0.0}

\title{Rounding Fermionic Moment Relaxations: Improved Gaussian Approximations for Hamiltonians with Classical Interactions}
\author[1]{Sre\'{c}ko \DJ{}ura\v{s}inovi\'{c}%
  \thanks{\texttt{srecko.durasinovic@monash.edu}}}
\author[2]{Hamza Fawzi%
  \thanks{\texttt{hf323@cam.ac.uk}}}
\author[1]{James Saunderson%
  \thanks{\texttt{james.saunderson@monash.edu}}}

\affil[1]{Department of Electrical and Computer Systems Engineering, Monash University, Australia}
\affil[2]{Department of Applied Mathematics and Theoretical Physics,
University of Cambridge, United Kingdom}

\begin{document}
\maketitle

\begin{abstract}
Fermionic Hamiltonians with density--density interactions arise, for example, 
in extended Hubbard models and real-space
representations of electronic-structure problems.
We study the approximation of their maximum energy by Slater
determinants. Our algorithm rounds a level-two fermionic
semidefinite moment relaxation into two candidate Slater
determinants and returns the one with larger energy, addressing
the semidefinite programming (SDP) rounding question raised in~\cite{STH}. For two models with positive
semidefinite terms, namely general classically interacting fermionic
Hamiltonians and Fermionic Max Cut, our proposed rounding scheme achieves expected constant
approximation ratios of $0.82699$ and $0.67971$, respectively.
We further analyze the rounding procedures for traceless
Hamiltonians on general graphs and obtain improved
constant-factor guarantees on bipartite interaction graphs. Finally, we show that augmenting our construction with a
postprocessing step that enforces an \textit{exact} particle number
yields a $\frac12$-approximation on bipartite interaction
graphs.
\end{abstract}
\tableofcontents

\section{Introduction}
\label{sec: intro}
A quantum system of $n$ fermionic modes, representing, for example,
spin-orbitals in a molecule or in a lattice model \cite{Hubbard1963,Babbush2018}, has $2^n$ occupation-basis states, since each mode can
be empty or occupied. Hence,  describing a general quantum state requires knowing a density matrix of size $2^n\times2^n$. However, when working with fermionic systems 
it is natural to restrict the study to \textit{fermionic Gaussian states}, which, in contrast,
are fully determined by a $2n\times2n$ covariance matrix.
For these states, Wick's theorem expresses higher-order
correlations in terms of the quadratic correlations recorded
in the covariance matrix, allowing their energies to be
evaluated efficiently~\cite{Wick,Bravyi,SuraceTagliacozzo}

Yet, being able to efficiently evaluate the energy of a given Gaussian state does not tell us how to find one with high energy.
This raises a central question studied in this paper:
\begin{quotation}
    How can one efficiently construct a
Gaussian state with a provable approximation guarantee,
i.e., with expected energy at least a constant fraction of
the unrestricted optimum on every instance of the
problem class (Definition~\ref{def:approximation-ratio})?
\end{quotation}
\begin{comment}
    This raises two different, but related questions:\begin{itemize}
    \item How much of the maximum energy
can a Gaussian state attain?
\item How can we efficiently construct a
Gaussian state with a provable approximation guarantee,
i.e., with expected energy at least a constant fraction of
the unrestricted optimum on every instance of the
problem class (Definition~\ref{def:approximation-ratio})?
\end{itemize}
\end{comment}
We study this question for \emph{classically interacting
fermionic Hamiltonians}, following~\cite{STH}. On $n$ modes,
these take the following general form
\begin{equation}\label{eq: H-intro}
     H=
    -\sum_{\{j,k\}\in E'}w'_{jk}
     \bigl(a_j^\dagger a_k+a_k^\dagger a_j\bigr)-\sum_{\{j,k\}\in E}w_{jk}n_jn_k+\sum_{j\in V}\mu_j n_j+\gamma \mathbb{I},
\end{equation}
where $V=\{1,\ldots,n\}$, $a_j^\dagger$ and $a_j$ create and
annihilate a fermion in mode $j$ (see equation \eqref{eq:creation-annihilation} for an explicit description of these operators), and $n_j=a_j^\dagger a_j$
records its occupation. Here $w_{jk}',\mu_j,\gamma\in\mathbb R$, $w_{jk}\ge0$ and $\mathbb{I}$ is the identity operator
on the $2^n$-dimensional fermionic state space (we also use this
notation for identity matrices of other sizes, with the
dimension understood from context).

The density--density interaction terms $-w_{jk}n_jn_k$,
indexed by the edges of $E$, penalize simultaneous occupation
of modes $j$ and $k$. These interactions are called \emph{classical} because their
energy on an occupation-basis state $|x\rangle$ is the
classical quadratic function
$-\sum_{\{j,k\}\in E}w_{jk}x_jx_k$ of the binary occupations
$x_j\in\{0,1\}$. Equivalently, the interaction operator is
diagonal in the occupation basis. The hopping terms
$-w'_{jk}(a_j^\dagger a_k+a_k^\dagger a_j)$, indexed by the
edges of $E'$, transfer fermions between modes $j$ and $k$.
Unlike the diagonal interactions, they couple different
occupation configurations, which is the source of the genuinely quantum behavior of the model.

Our goal is to construct a Slater determinant whose expected
energy approximates $\OPT(H):=\lambda_{\max}(H)$,
where $\lambda_{\max}(H)$ denotes the largest eigenvalue
of $H$, equivalently, its maximum energy over all
quantum states.  Slater determinants are pure fermionic Gaussian states
obtained by occupying a fixed set of orthonormal modes. Computing $\OPT(H)$ to inverse-polynomial additive precision
is QMA-hard, by an adaptation of the electronic-structure
hardness result of~\cite{SV,OIWF}. 
We seek instead polynomial-time \textit{Gaussian approximation
algorithms}, i.e., algorithms that construct Gaussian states
whose expected energy is at least a \textit{constant} fraction
$\alpha\in(0,1]$ of $\OPT(H)$. The choice of $\gamma$ in \eqref{eq: H-intro}
matters because shifting all energies by the same constant
does not change which states maximize the energy, but it
does change what fraction of the optimum a given state
achieves. We consider the same normalizations as in~\cite{STH} and
derive approximation guarantees for the corresponding
families of Hamiltonians.
%%%%%%%%%%%%%%%%%%%%%%%%%%%%%%%%%%%%%%%%%%%%%%%%%%%%%%%%%%%%%%%%%%%%%%%%%%%%%%%%
\paragraph{Gaussian-based approximation algorithms.}
For general traceless Hamiltonians with quadratic and quartic
fermionic terms, \cite{BGKT}
gave an efficient Gaussian approximation algorithm with ratio
$\Omega((n\log n)^{-1})$. Their method uses Wick's theorem to
express optimization over Gaussian states as a quadratic
optimization problem in the covariance matrix, to which they
apply semidefinite relaxation and rounding.
Subsequent work showed that
constant Gaussian-based guarantees are impossible for arbitrary quartic interactions. Namely, for the dense SYK-$4$ model \cite{Kitaev}, the best Gaussian energy is only an
$\mathcal{O}(n^{-\frac12})$ fraction of the optimum, as established in the
study of strongly interacting fermions \cite{HastingsODonnell} and extended to higher-degree
SYK models in \cite{HSHT}. Thus, constant Gaussian approximation guarantees require
additional structure in the Hamiltonian.

Sparsity provides one example of such structure. For sparse Hamiltonians with quadratic and quartic Majorana
terms, \cite{HSHT} obtained an efficient
Gaussian approximation with factor $\Omega(k^{-2})$ whenever
each Majorana operator occurs in at most $k$ nonzero terms.
\cite{HPT} further improved this dependence
to $\Omega(k^{-1})$. Their construction exploits sparsity to select a collection
of Hamiltonian terms carrying a substantial fraction of the
total absolute coefficient weight, and produces a Gaussian
state whose energy is at least the sum of the absolute
coefficients of the selected terms.\\
More recently, \cite{Parekh2026}
proved a classical $\Omega(n^{-\frac12})$-approximation for general
traceless quartic Hamiltonians, by considering an algorithm that compares a
Gaussian candidate from \cite{BGKT} with a superposition of polynomially many
occupation-basis states.

Alternatively, one can exploit the form of the interactions
rather than sparsity. For Hamiltonians with classical
interactions, \cite{STH} obtained constant Gaussian
approximation guarantees even when the interaction graph
is dense. Their main tool, called \emph{Gaussian blend}, 
combines covariance matrices of Gaussian states to produce
another Gaussian state, retaining part of the hopping
energy while controlling the loss from the classical
interactions. They proved the existence of a fermionic Gaussian state achieving a \(\frac13\)-approximation ratio for the traceless model and provided polynomial-time constructions for bipartite interaction graphs and models with positive semidefinite terms, including their Fermionic Max Cut model. 
Since~\cite{STH} is the work most closely related
to ours, Table~\ref{tab:guarantees} compares its
approximation guarantees with those obtained
in this paper.

\paragraph{Rounding moment relaxations.} Fermionic moment relaxations upper-bound the maximum
energy by replacing the requirement that expectation
values arise from a physical quantum state with a
set of necessary conditions, including positivity
and the identities implied by the canonical
anticommutation relations of fermionic creation
and annihilation operators. 
These conditions are imposed only up to a prescribed
degree, giving a hierarchy of increasingly large
semidefinite programs (SDPs) with progressively tighter
upper bounds~\cite{PNA}. 
At level two, the relaxation includes quadratic and
quartic pseudo-moments, which do not need to satisfy the Wick
identities obeyed by Gaussian states. The challenge is
then to round the SDP solution to a physical state
whose energy retains a guaranteed fraction of the
relaxation bound. For the fermionic models considered
here, how to achieve such rounding with Gaussian states
was left open in~\cite{STH}.

For classical Max Cut, \cite{GW} showed how to round the
vector solution of a semidefinite relaxation to a cut
using a random hyperplane. For higher 
levels of SDP hierarchies, \cite{BRS}
developed methods that condition on a subset of variables
and control the remaining correlations to bound the rounding
loss. Additional difficulty arises when the rounded assignment is required to satisfy some global constraints, such as a prescribed
cardinality. One example is the hierarchy rounding for Max Bisection
in~\cite{RaghavendraTan}, where the two sides of the cut
must have equal size.

For quantum Hamiltonians, SDP rounding similarly aims to
construct physical states from relaxed operator correlations
with provable energy guarantees.
Using level-two positivity constraints, including monogamy
inequalities, \cite{PTLevelTwo,PTProduct} established a tight
$\frac12$-approximation ratio using product states for
Quantum Max Cut, introduced in~\cite{GharibanParekhQMC}.
Subsequently, \cite{LeeParekh} improved the ratio to
$0.595$ by combining SDP rounding with matching-based
states on disjoint edges.
We emphasize that Quantum Max Cut differs from the
Fermionic Max Cut problem studied in
Subsection~\ref{sec:fmc}, although the two are
equivalent on a line via the Jordan--Wigner transformation \cite{JordanWigner1928}. Beyond Quantum Max Cut, SDP rounding has also been
studied for dense local Hamiltonians~\cite{BrandaoHarrow2016}
and general two-local Hamiltonians with positive
semidefinite terms~\cite{HallgrenLeeParekh2020}.
For fermionic Hamiltonians with classical interactions, \cite[Appendix~D]{STH} suggested
that rounding fermionic moment relaxations to Gaussian states while preserving a constant fraction
of the relaxation bound could potentially improve the approximation ratio. This is precisely the path we follow in this paper.
%%%%%%%%%%%%%%%%%%%%%%%%%%%%%%%%%%%%%%%%%%%%%%%%%%%%%%%%%%%%%%
\paragraph{Our contribution}
We propose two rounding schemes that each construct a Slater
determinant from a solution of the level-two fermionic moment
relaxation of the original problem, and return the candidate with larger energy.
Our analysis focuses on their expected energies on each
interaction edge and controls the rounding loss using
inequalities implied by the level-two positivity constraints
satisfied by the relaxation solution. For different Hamiltonian families and different normalizations, we summarize our obtained results in 
Table~\ref{tab:guarantees}.
\begin{table}[H]
\centering
\small
\renewcommand{\arraystretch}{1.25}
\begin{tabular}{@{}>{\raggedright\arraybackslash}p{0.33\linewidth}>{\raggedright\arraybackslash}p{0.30\linewidth}>{\raggedright\arraybackslash}p{0.30\linewidth}@{}}
\toprule
\textbf{Model and assumptions} & \textbf{Previous Gaussian-based approximation guarantee} & \textbf{This work} \\
\midrule
Positive semidefinite terms
& $0.637$~\cite[Theorem~2.5]{STH}
& $\bf0.82699$ (Theorem~\ref{thm:PSD-guarantee}) \\
Fermionic Max Cut, \eqref{eq:Hfmc}
& $\frac12$~\cite[Proposition~5.6]{STH}
& $\bf0.67971$ (Theorem~\ref{thm:fmc}) \\
Traceless, general graphs, \eqref{eq:H-traceless}
& $\frac13$, existence only~\cite[Theorem~2.2]{STH}
& Additive bound only\textsuperscript{a} 
(Theorem~\ref{thm:traceless-additive}) \\
Traceless, bipartite graphs
& $\frac13$~\cite[Theorem~2.3]{STH}
& $\bf\frac12$ (Theorem~\ref{thm:traceless-bipartite}) \\
Traceless, bipartite graphs, mean $q$-particle number\textsuperscript{b}, \eqref{eq:particle-number-optimum}
& $\frac{1}{5-4\frac qn}$~\cite[Theorem~2.7]{STH}
& $\bf\frac12$ (Theorem~\ref{thm:particle-number-bipartite}) \\
Traceless, bipartite graphs, \textit{exact} $q$-particle number,  \eqref{eq:exact-q-opt}
& No previous guarantee
& $\bf\frac12$ (Corollary~\ref{cor:exact-particle-bipartite}) \\
\bottomrule
\end{tabular}
\vspace{0.25em}
\begin{minipage}{0.96\linewidth}
\footnotesize
\raggedright
\setlength{\parindent}{0pt}
\setlength{\parskip}{0.3em}
\textsuperscript{a} Additive bounds control the difference, not the ratio, 
between the SDP upper bound and the expected energy
of the rounded state.
\textsuperscript{b} Here 
$ q\in\left\{0,\dots,\left\lfloor \frac n2\right\rfloor\right\}$.
\end{minipage}
\caption{Overview of Gaussian approximation guarantees
for the Hamiltonian models studied in~\cite{STH}
and further investigated in this work.
Numerical entries in the second and third columns
denote multiplicative approximation guarantees.
Bipartiteness is required only for the interaction
graph $(V,E)$, while the hopping graph $(V,E')$
is unrestricted.}
\label{tab:guarantees}
\end{table}
The two schemes use different parts of the relaxation
solution. The first uses the spectral
decomposition of the one-particle correlation matrix
to construct a Slater determinant by independently
occupying the resulting modes according to the
corresponding eigenvalues. This preserves all quadratic
correlations in expectation, while Wick's theorem allows
us to compute the change in the interaction energy
(Subsection~\ref{sec:quadratic-rounding}). The second proposed procedure (see Subsection \ref{sec:occupation-rounding}) uses random-hyperplane rounding
of the occupation correlations to produce an occupation-basis
Slater determinant, whose hopping energy is zero. The approximation guarantees are then established by bounding
the expected energies of the two rounding schemes relative
to the optimal relaxation value.

In the presence of particle-number constraints (see Subsection~\ref{sub: particle-number-constraints}), our first rounding scheme alone yields a
Gaussian state with \textit{mean} particle number
$q\in\left\{0,\dots,\left\lfloor \frac n2\right\rfloor\right\}$. We then perform an iterative postprocessing procedure
that modifies two occupation probabilities at a time,
while preserving their sum and without decreasing the energy. After finitely many iterations, we obtain a Slater determinant
with \textit{exactly} $q$ particles and the same
$\frac12$-approximation guarantee as for the
mean-particle-number problem on bipartite interaction
graphs, providing a solution to the fixed-$q$-particle-number
question left open in~\cite{STH}.

\medskip
The rest of the paper is structured as follows. Section \ref{sec: prelim} introduces Gaussian states and the fermionic moment
hierarchy. Section~\ref{sec:rounding} presents the two complementary rounding schemes. Sections~\ref{sec:PSD-guarantees}
and~\ref{sec:traceless-guarantees} establish the guarantees
for the positive semidefinite and traceless models, including the case with particle-number constraints. Finally, Section~\ref{sec:discussion} discusses limitations and further
questions.

%%%%%%%%%%%%%%%%%%%%%%%%%%%%%%%%%%%%%%%%%%%%%%%%%%%%%%%%%%%%%%%%%%%%%%%%%%%%%%%%%%%%%%%%%%%%%%%%%%%%%%%%%%%%%%%%%%5
\section{Preliminaries}
\label{sec: prelim}
\subsection{Basic fermionic algebra}
We consider a system of $n$ fermionic modes, indexed by
$V=\{1,\ldots,n\}$ (see, e.g., \cite{NegeleOrland1988} for all the terminology and language in this section). The corresponding fermionic Fock space
$\mathcal F_n\cong\mathbb C^{2^n}$ has an orthonormal
occupation-number basis $\{|x\rangle=|x_1\cdots x_n\rangle\mid x_j\in\{0,1\}, j\in V\}$, 
where $x_j=0$ means that mode $j$ is empty and $x_j=1$
means that it is occupied. The annihilation and creation operators
$\{a_j,a_j^\dagger\}_{j\in V}$ act on this basis as follows:
\begin{equation}\label{eq:creation-annihilation}
\begin{aligned}
a_j|x_1\cdots x_j\cdots x_n\rangle
&=
\begin{cases}
0, & x_j=0,\\
(-1)^{\sum_{k<j}x_k}
|x_1\cdots0\cdots x_n\rangle, & x_j=1,
\end{cases}
\\[1ex]
a_j^\dagger|x_1\cdots x_j\cdots x_n\rangle
&=
\begin{cases}
(-1)^{\sum_{k<j}x_k}
|x_1\cdots 1\cdots x_n\rangle, & x_j=0,\\
0, & x_j=1.
\end{cases}
\end{aligned}
\end{equation}
These operators satisfy
the canonical anticommutation relations (CAR):
\begin{align}\label{eq:CAR}
a_j^\dagger a_k+a_ka_j^\dagger
=\delta_{jk}\mathbb I, \quad
a_ja_k+a_ka_j=0,\quad
a_j^\dagger a_k^\dagger+a_k^\dagger a_j^\dagger
=0,
\end{align}
where $\delta_{jk}$ is the Kronecker delta.
In particular, $(a_j^\dagger)^2=0$, expressing
the fact that a fermionic mode cannot be occupied twice. The vacuum state $|\mathrm{vac}\rangle=|0,\ldots,0\rangle$
satisfies $a_j|\mathrm{vac}\rangle=0$ for every $j\in V$.
Every occupation-number basis state therefore can be obtained by
applying creation operators to the vacuum:
\[
|x\rangle
=(a_1^\dagger)^{x_1}\cdots
(a_n^\dagger)^{x_n}|\mathrm{vac}\rangle,
\quad x\in\{0,1\}^n.
\]
The occupation of mode $j$ is described by the
number operator
\[
n_j:=a_j^\dagger a_j,
\qquad
n_j|x\rangle=x_j|x\rangle.
\]
Thus, $n_j$ has eigenvalues $0$ and $1$, and satisfies
$n_j^2=n_j$ and $n_j^\dagger=n_j$. Hence, $n_j$ is
the orthogonal projector onto the subspace in which
mode $j$ is occupied.
%%%%%%%%%%%%%%%%%%%%%%%%%%%%%%%%%%%%%%%%%%%%%%%%%%%%%%%%%%%%%
\subsection{Hamiltonians and the optimization problem}
A quantum state on $\mathcal F_n$ is represented by a density
matrix $\rho\in\mathbb C^{2^n\times2^n}$ satisfying
$\rho\succeq0$ and $\operatorname{Tr}(\rho)=1$. It is \emph{pure} if $\rho=|\psi\rangle\langle\psi|$ for a
unit vector $|\psi\rangle\in\mathcal F_n$; otherwise, it is
\emph{mixed}. For a Hermitian operator
$A:\mathcal F_n\to\mathcal F_n$, we write
\[
    \langle A\rangle_\rho:=\operatorname{Tr}(\rho A).
\]
In particular, $\langle n_j\rangle_\rho$ is the probability
that mode $j\in V$ is occupied in the state $\rho$.

A Hamiltonian for an $n$-mode fermionic system is a
Hermitian operator $H:\mathcal F_n\to\mathcal F_n$,
with $\langle H\rangle_\rho$ denoting its mean energy
when the system is in the state $\rho$. We study Hamiltonians consisting of hopping terms, repulsive pairwise occupation interactions, and  occupation terms. Their generic form corresponds therefore to:
\begin{equation}\label{eq: H-generic}
    H=
    -\sum_{\{j,k\}\in E'}w'_{jk}
     \bigl(a_j^\dagger a_k+a_k^\dagger a_j\bigr)-\sum_{\{j,k\}\in E}w_{jk}n_jn_k+\sum_{j\in V}\mu_j n_j+\gamma \mathbb{I},
\end{equation}
where $\mathbb{I}$ represents the identity matrix,  $w'_{jk}\in\mathbb{R}$ for all $\{j,k\}\in E'$, $w_{jk}\ge0$ for all $\{j,k\}\in E$, $\mu_j\in\mathbb R$ for all $j\in V$, and $\gamma\in\mathbb R$.
The edge sets $E'$ and $E$ specify the hopping terms and the occupation
interactions, respectively. These occupation interactions are called \emph{classical} because
optimizing their energy alone is equivalent to optimizing
the quadratic function $-\sum_{\{j,k\}\in E}w_{jk}x_jx_k$
over binary occupations $x\in\{0,1\}^n$. Indeed,
$n_jn_k|x\rangle=x_jx_k|x\rangle$, so the interaction
operator is diagonal in the occupation-number basis. Notice that requiring the weights $\{w_{jk}\}_{\{j,k\}\in E}$ to be nonnegative penalizes simultaneous occupation of different modes
$j$ and $k$. 
The hopping terms, by contrast, transfer fermions between
modes and couple different occupation-number basis states.
Consequently, the full Hamiltonian is generally not
diagonal in this basis, and its energy cannot be determined
from occupation probabilities alone.

Throughout the paper, we maximize energy of Hamiltonians in question. Therefore, our optimization problem can be written as follows:
\begin{align}\label{eq: OPTProb}
    \operatorname{OPT}(H)
    :=
    \max_{\substack{
        \rho\in{\mathbb C}^{2^n\times2^n},\ \rho\succeq0,\\
        \operatorname{Tr}(\rho)=1
    }}
    \operatorname{Tr}(\rho H)
    = \max_{\substack{
        |\psi\rangle\in\mathcal F_n,\\
        \langle\psi|\psi\rangle=1
    }}
    \langle\psi|H|\psi\rangle = 
    \lambda_{\max}(H),
\end{align}
where $ \lambda_{\max}$ denotes largest-eigenvalue
function on Hermitian operators.  Our goal is to construct a state whose energy is at least
a constant fraction of $\operatorname{OPT}(H)$, uniformly
over all instances of the Hamiltonian family under
consideration. For randomized algorithms, this fraction
is measured in terms of the expected energy of the
output state. We formalize this guarantee as follows.

\begin{definition}[Expected $\alpha$-approximation ratio]
\label{def:approximation-ratio}
Let $\mathcal H$ be a family of Hamiltonians with
$\operatorname{OPT}(H)>0$ for every $H\in\mathcal H$.
For $\alpha\in]0,1]$, an algorithm is an \emph{expected
$\alpha$-approximation} if, for every input
$H\in\mathcal H$, it returns a quantum state $\rho_{\mathrm{alg}}$
satisfying
\begin{equation}
\label{eq:approximation-ratio}
    \mathbb E\!\left[\langle H\rangle_{\rho_{\mathrm{alg}}}\right]
    \ge \alpha\,\operatorname{OPT}(H).
\end{equation}
\end{definition}
In this work, we seek (efficient) algorithms that achieve the
approximation guarantee in Definition~\ref{def:approximation-ratio} using fermionic Gaussian states, whose properties we
recall next.

\subsection{Fermionic Gaussian states}
\label{sec:gaussian}
Fermionic Gaussian states form a family of quantum states
that can be described using only quadratic correlations.
This makes them useful for approximating interacting
fermionic systems, since,  for a given Hamiltonian $H$, the energy
$\langle H\rangle_\rho$ of a Gaussian state $\rho$ can
be evaluated efficiently, 
without storing an exponentially large density matrix. Following \cite{Bravyi,SuraceTagliacozzo}, we recall their definition and basic properties.

\begin{definition}[Fermionic Gaussian state]
\label{def:gaussian}
Let $\mathcal Q_n$ be the set of Hermitian quadratic fermionic operators
of the form
\begin{equation}
\label{eq:operatorK}
\begin{aligned}
K={}&\sum_{j,k=1}^n A_{jk}a_j^\dagger a_k+\sum_{1\le j<k\le n}
\left(B_{jk}a_j^\dagger a_k^\dagger
      +\overline{B_{jk}}a_ka_j\right),
\quad A=A^\dagger,\quad B_{jk}\in\mathbb C.
\end{aligned}
\end{equation}
A density matrix $\rho$ on $\mathcal F_n$ describes a \emph{fermionic Gaussian state} if there exists a sequence $(K_m)_{m\ge1}$ in $\mathcal Q_n$ such that
\begin{equation}
\label{eq:gaussian-closure}
\lim_{m\to\infty}\frac{e^{-K_m}}{\operatorname{Tr}(e^{-K_m})}
=\rho.
\end{equation}
\end{definition}
The definition includes every state of the form
$\rho=\frac{e^{-K}}{\operatorname{Tr}(e^{-K})}$ with $K\in\mathcal Q_n$,
by choosing $K_m=K$ for all $m$.
Since $e^{-K}$ is positive definite, these states have full rank
and are therefore mixed. Pure Gaussian states are obtained as limits of these normalized matrix exponentials. Finally, additive constants from~\eqref{eq:operatorK} are omitted since, for $c\in\R$,
replacing $K$ by $K+c\mathbb{I}$ leaves the normalized density matrix unchanged. 

The operators in the second sum of~\eqref{eq:operatorK}
create or annihilate pairs of fermions. Hence, general Gaussian states need not necessarily conserve the total particle number.
However, the states that we will construct in this paper will belong to the following subclass.
\begin{definition}[Number-conserving Gaussian state]
\label{def:number-conserving-gaussian}
A (Gaussian) state $\rho$ is \emph{number-conserving} if
\begin{equation}
\label{eq:number-conserving}
    \rho N=N\rho,
\end{equation}
where $N:=\sum_{j=1}^n n_j$ is the total particle-number operator.
\end{definition}
We construct such states by setting $B_{jk}=0$
in~\eqref{eq:operatorK}, leaving only terms of the form
$a_j^\dagger a_k$. These operators preserve particle
number, so $KN=NK$. The associated normalized exponential
therefore commutes with $N$, as do limits of these states. We emphasize that a state does not need to have a fixed
number of particles to be number conserving. Indeed,
it may be a mixture of states with different fixed
particle numbers. In general, a state $\rho$ that is number-conserving is block diagonal
in the occupation basis, with each block corresponding
to a fixed particle number. 

Since $a_ja_k$ and
$a_j^\dagger a_k^\dagger$ decrease and increase
the particle number by two, respectively, it is not hard to show that their expectation values vanish in any
number-conserving (Gaussian) state:
\begin{equation}
\label{eq:pairing-zero}
    \langle a_ja_k\rangle_\rho
    =\langle a_j^\dagger a_k^\dagger\rangle_\rho=0.
\end{equation}
The remaining quadratic expectations involving one creation and
one annihilation operator are described by the
\textit{one-particle correlation matrix} $R\in\mathbb C^{n\times n}$,
defined entry-wise by
\begin{equation}
\label{eq:one-particle-matrix}
R_{jk}:=\langle a_j^\dagger a_k\rangle_\rho,
\quad j,k\in V.
\end{equation}
By construction,
\begin{equation}
\label{eq:particle-trace}
\operatorname{Tr}(R)=\sum_{j\in V}\langle n_j\rangle_\rho
=\langle N\rangle_\rho.
\end{equation}
Moreover, off-diagonal entries of $R$ determine the hopping expectations in
\eqref{eq: H-generic}, namely:
\begin{equation}
\label{eq:hopping-from-R}
\langle a_j^\dagger a_k+a_k^\dagger a_j\rangle_\rho
=R_{jk}+R_{kj}=2\operatorname{Re}(R_{jk}).
\end{equation}
Let us recall some properties that one-particle matrices satisfy.
\begin{proposition}
\label{prop:one-particle-bounds}
    For every density matrix $\rho$ (not necessarily describing a Gaussian state), the one-particle correlation matrix defined
in \eqref{eq:one-particle-matrix} satisfies
\begin{equation}
\label{eq:one-particle-bounds}
R=R^\dagger\quad\text{and}\quad 0\preceq R\preceq \mathbb{I}.
\end{equation}
\end{proposition}
\begin{proof}
The matrix $R$ is Hermitian, since for all $j,k\in V$, 
$\overline{R_{jk}}
    =\overline{\operatorname{Tr}(\rho a_j^\dagger a_k)}
    =\operatorname{Tr}(\rho a_k^\dagger a_j)
    =R_{kj}$. 
For $z\in\mathbb C^n$, set $b:=\sum_jz_ja_j$.
Since $\rho$ is positive semidefinite,
\[
z^\dagger Rz=\operatorname{Tr}(\rho b^\dagger b)\ge0.
\]
The CAR from \eqref{eq:CAR} give $bb^\dagger+b^\dagger b=\|z\|_2^2\mathbb{I}$, and hence
\[
z^\dagger(\mathbb{I}-R)z
=\|z\|_2^2\operatorname{Tr}(\rho)-\operatorname{Tr}(\rho b^\dagger b)
=\operatorname{Tr}(\rho bb^\dagger)\ge0.
\]
Both inequalities hold for every $z$, thereby proving the stated result.
\end{proof}
For Gaussian states, Wick's theorem determines higher-order expectations
from quadratic ones~\cite{Wick,Molinari,SuraceTagliacozzo}. Together
with~\eqref{eq:pairing-zero}, it gives the following identity, which we
will use to evaluate occupation interactions.

\begin{proposition}
\label{prop:wick-occupations}
Let $\rho$ be a number-conserving Gaussian state with one-particle
matrix $R$. For distinct modes $j,k\in V$,
\begin{equation}
\label{eq:wick-occupations}
\langle n_jn_k\rangle_\rho=R_{jj}R_{kk}-|R_{jk}|^2.
\end{equation}
\end{proposition}
\begin{proof}
The four-operator form of Wick's theorem \cite[Equations (48)-(49), page 13]{SuraceTagliacozzo} gives
\begin{align*}
\langle n_jn_k\rangle_\rho=\langle a_j^\dagger a_j a_k^\dagger a_k\rangle_\rho
={}&\langle a_j^\dagger a_j\rangle_\rho
    \langle a_k^\dagger a_k\rangle_\rho-\langle a_j^\dagger a_k^\dagger\rangle_\rho
    \langle a_j a_k\rangle_\rho+\langle a_j^\dagger a_k\rangle_\rho
    \langle a_j a_k^\dagger\rangle_\rho.
\end{align*}
The second term vanishes by~\eqref{eq:pairing-zero}.
For $j\ne k$, the CAR give
$\langle a_j a_k^\dagger\rangle_\rho=-R_{kj}$.
The remaining terms are therefore
$R_{jj}R_{kk}-R_{jk}R_{kj}=R_{jj}R_{kk}-|R_{jk}|^2$.
\end{proof}
More generally, Wick's theorem fixes every higher-order correlation from
$R$ within the number-conserving Gaussian family. Without Gaussianity assumption, quantum states with the same $R$ can have different joint occupations.

For a number-conserving Gaussian state $\rho$,
Proposition~\ref{prop:wick-occupations} allows us to  evaluate the energy of the Hamiltonian \eqref{eq: H-generic} using the $n\times n$ matrix $R$ alone,
without constructing the full $2^n\times 2^n$ density matrix, namely:
\begin{equation}
\label{eq:gaussian-energy}
\begin{aligned}
\langle H\rangle_\rho
={}&\gamma+\sum_{j\in V}\mu_jR_{jj}
-\sum_{\{j,k\}\in E}w_{jk}
  \bigl(R_{jj}R_{kk}-|R_{jk}|^2\bigr)-2\sum_{\{j,k\}\in E'}w'_{jk}\operatorname{Re}(R_{jk}).
\end{aligned}
\end{equation}
So far, we have considered a Gaussian state $\rho$ and
expressed its energy in terms of its one-particle
correlation matrix $R$. Our first rounding scheme proceeds
in the opposite direction: starting from a matrix $R$
obtained from the SDP relaxation, we construct a Gaussian
state $\rho_R$ with the same one-particle correlations.
The following proposition guarantees that such a
construction is possible~\cite[Sections~3.2--3.6]{SuraceTagliacozzo}.
\begin{proposition}
\label{prop:gaussian-realization}
For every Hermitian matrix $R\in\mathbb C^{n\times n}$ with
$0\preceq R\preceq \mathbb{I}$, there is a unique number-conserving Gaussian
state $\rho_R$ satisfying
$\langle a_j^\dagger a_k\rangle_{\rho_R}=R_{jk}$ for all $j,k\in V$.
\end{proposition}
\begin{proof}
First, the matrix $R$ being Hermitian, it admits the following  eigenvalue decomposition
\begin{equation}
\label{eq:R-eigendecomposition}
R=U\operatorname{diag}(\nu_1,\ldots,\nu_n)U^\dagger,
\quad U^\dagger U=UU^\dagger=\mathbb{I},\quad  0\le\nu_\ell\le1.
\end{equation}
Using the columns of $U$, define new fermionic annihilation
and creation operators by
\begin{equation}
\label{eq:normal-modes}
b_\ell:=\sum_{j=1}^n U_{j\ell}a_j,
\quad
b_\ell^\dagger:=\sum_{j=1}^n
\overline{U_{j\ell}}a_j^\dagger,\quad l\in V.
\end{equation}
Unitarity of $U$ implies that these operators satisfy the CAR from \eqref{eq:CAR}. To prove the result, we construct $\rho_R$ by occupying
each mode associated with $b_\ell$ independently with
probability $\nu_\ell$, and verify that its associated one-particle matrix is precisely $R$.

For each $x\in\{0,1\}^n$, let
\begin{equation}
\label{eq:mode-occupation-state}
|\psi_{x}\rangle:=(b_1^\dagger)^{x_1}\cdots
                    (b_n^\dagger)^{x_n}|\mathrm{vac}\rangle.
\end{equation}
This state occupies mode $\ell\in V$ exactly when $x_\ell=1$. Hence, $\{|\psi_x\rangle\}_{x\in\{0,1\}^n}$ is an orthonormal occupation basis of $\mathcal F_n$
associated with the operators $b_1,\ldots,b_n$.
Let us now define
\begin{equation}
\label{eq:gaussian-realization}
\rho_R:=\sum_{{x}\in\{0,1\}^n}p_R({x})
        |\psi_{x}\rangle\langle\psi_{x}|,
\quad\text{with}\quad
p_R({x}):=\prod_{l\in V}\nu_\ell^{x_l}(1-\nu_\ell)^{1-x_l}.
\end{equation}
It follows that, for all $x,y\in\{0,1\}^n$, $\langle\psi_x|\rho_R|\psi_y\rangle
=p_R(x)\delta_{xy}$, i.e., the matrix \(\rho_R\) is diagonal in the basis $\{|\psi_x\rangle\}_{x\in\{0,1\}^n}$, with diagonal entries equal to \(p_R(x)\). Since $p_R$ is the joint probability distribution of
independent Bernoulli variables with parameters
$\nu_1,\ldots,\nu_n$, we also have
$\sum_{x\in\{0,1\}^n}p_R(x)=1$. Thus, $\rho_R$ is a probabilistic mixture of normalized
pure states and is therefore a density matrix.

By construction, mode $\ell\in V$ is occupied with probability
$\nu_\ell\in[0,1]$, so
$\langle b_\ell^\dagger b_\ell\rangle_{\rho_R}=\E[x_l]=\nu_\ell$.
For $\ell\ne m$, the operator $b_\ell^\dagger b_m$ either maps $|\psi_x\rangle$ to zero or transfers a fermion from mode $m$ to
mode $\ell$, producing an orthogonal occupation state.
Hence, $\langle b_\ell^\dagger b_m\rangle_{\rho_R}
=\sum_{x\in\{0,1\}^n} p_R(x)
  \langle\psi_x|b_\ell^\dagger b_m|\psi_x\rangle
=\nu_\ell\delta_{\ell m}$. Using~\eqref{eq:normal-modes}, we therefore obtain
\[
\langle a_j^\dagger a_k\rangle_{\rho_R}
=\sum_{\ell,m\in V}
U_{j\ell}\overline{U_{km}}\,
\langle b_\ell^\dagger b_m\rangle_{\rho_R}
=\sum_{l\in V} U_{j\ell}\nu_\ell\overline{U_{k\ell}}
=R_{jk},
\]
where the last equality follows
from~\eqref{eq:R-eigendecomposition}. We will now prove, by distinguishing two possible cases, that this density matrix is Gaussian.

\textit{Case I: $\forall l\in V,\; 0<\nu_\ell<1$.} Define the following operator:
\begin{equation}
\label{eq:gaussian-realization-generator}
K_R:=\sum_{\ell=1}^n
\log\!\left(\frac{1-\nu_\ell}{\nu_\ell}\right)
b_\ell^\dagger b_\ell.
\end{equation}
By definition, this is a Hermitian quadratic operator of the form \eqref{eq:operatorK}.
Moreover, since $b_\ell^\dagger b_\ell|\psi_{x}\rangle={x}_\ell|\psi_{x}\rangle$, $K_R$ is diagonal in the occupation basis, and satisfies
\begin{align*}
e^{-K_R}|\psi_{x}\rangle
&=e^{-\sum_{\ell=1}^n
x_\ell\log\left(\frac{1-\nu_\ell}{\nu_\ell}\right)}
|\psi_x\rangle=\prod_{\ell=1}^n
  \left(\frac{\nu_\ell}{1-\nu_\ell}\right)^{x_\ell}
  |\psi_{x}\rangle,\\
\operatorname{Tr}(e^{-K_R})
&=\sum_{{x}\in\{0,1\}^n}\prod_{\ell=1}^n
  \left(\frac{\nu_\ell}{1-\nu_\ell}\right)^{x_\ell}
 =\prod_{\ell=1}^n\left(1+\frac{\nu_\ell}{1-\nu_\ell}\right)
 =\prod_{\ell=1}^n\frac{1}{1-\nu_\ell}.
\end{align*}
Dividing $e^{-K_R}$ by its trace, the diagonal entry associated
with $|\psi_{x}\rangle$ becomes exactly
\begin{align}\nonumber
\left\langle\psi_{x}\left|
\frac{e^{-K_R}}{\operatorname{Tr}(e^{-K_R})}
\right|\psi_{x}\right\rangle
=
\prod_{\ell=1}^n
(1-\nu_\ell)
\left(\frac{\nu_\ell}{1-\nu_\ell}\right)^{x_\ell}
=p_R({x}).
\end{align}
Hence, both $\rho_R$ and $\frac{e^{-K_R}}{\operatorname{Tr}(e^{-K_R})}$ are diagonal in the basis $\{|\psi_{x}\rangle\}_{x\in\{0,1\}^n}$, with the same
diagonal entries $p_R({x})$. Therefore, $\rho_R=\frac{e^{-K_R}}{\operatorname{Tr}(e^{-K_R})}$
satisfies~\eqref{eq:gaussian-closure} with the constant
sequence $K_m=K_R$, and is thus a fermionic Gaussian state.

\textit{Case II: $\exists l\in V,\;\nu_l=0$ or $\nu_l=1$.} The logarithm in
\eqref{eq:gaussian-realization-generator} becomes undefined. The
probabilities in~\eqref{eq:gaussian-realization}, however, remain
well defined, i.e., the mode $l$ is certainly empty or occupied, respectively. One can therefore set, for $0<\varepsilon<1$, $\nu_l=\varepsilon$ or $\nu_l=1-\varepsilon$, leaving all other
eigenvalues unchanged. Thus, \textit{Case~I} now applies, yielding a Gaussian state. As $\varepsilon\to0$, its diagonal entry corresponding to
$|\psi_x\rangle$ converges to $p_R(x)$ for every
$x\in\{0,1\}^n$. The resulting density matrices therefore converge to
$\rho_R$, which is Gaussian by Definition~\ref{def:gaussian}.

It remains to verify number conservation.
Each $|\psi_{x}\rangle$ has exactly $\sum_{l\in V} x_\ell$ particles,
so $N|\psi_{x}\rangle
    =\left(\sum_{l\in V} x_\ell\right)|\psi_{x}\rangle$. Thus $N$ and $\rho_R$ are diagonal in the same basis
$\{|\psi_{x}\rangle\}_{x\in\{0,1\}^n}$ and therefore commute.
Hence, $\rho_R$ is number-conserving.

Finally, for uniqueness, observe that any two number-conserving Gaussian states with
the same matrix $R$ have identical quadratic expectations.
By Wick's theorem, they agree on all products of creation and
annihilation operators. Since these products span the operator
space on $\mathcal F_n$, their two associated density matrices must coincide.
\end{proof}

\begin{corollary}[Pure Gaussian states]
\label{cor:gaussian-purity}
The Gaussian state $\rho_R$ in
Proposition~\ref{prop:gaussian-realization}
is pure if and only if
\[
R^2=R.
\]
\end{corollary}

\begin{proof}
Recall that the eigenvalues of $\rho_R$ are exactly
the weights $p_R(x)$. Purity is equivalent to
$\operatorname{rank}(\rho_R)=1$, so $\rho_R$ is pure
exactly when one of its eigenvalues equals $1$
and all others vanish.

If there exists $\ell\in V$ satisfying
$0<\nu_\ell<1$, the product formula for $p_R(x)$
assigns positive weight to at least two patterns,
differing in the occupation of mode $\ell$.
Hence $\rho_R$ is not pure.
Conversely, if $\nu_\ell\in\{0,1\}$ for every
$\ell\in V$, the unique occupation pattern
$x=(\nu_1,\ldots,\nu_n)$ has weight $1$.

Therefore, $\rho_R$ is pure exactly when all
eigenvalues of $R$ belong to $\{0,1\}$,
i.e., when $R^2=R$.
\end{proof}
\begin{remark}[Slater determinants]\label{rem:Slater}
Pure number-conserving Gaussian states are called
\emph{Slater determinants}. In the setting of
Corollary~\ref{cor:gaussian-purity}, such a state
has the form $\rho_R=|\psi_x\rangle\langle\psi_x|$,
where $x=(\nu_1,\ldots,\nu_n)\in\{0,1\}^n$.
It occupies precisely the modes with $\nu_\ell=1$
and contains exactly
$\sum_{\ell\in V}\nu_\ell=\operatorname{rank}(R)$
particles. In particular, every occupation-number
basis state $|x\rangle$, including the vacuum,
is a Slater determinant.
\end{remark}
\subsection{Fermionic moment relaxations}
\label{sec:hierarchy}
Proposition~\ref{prop:gaussian-realization} shows how to construct a
number-conserving Gaussian state from any Hermitian matrix
$R$ satisfying $0\preceq R\preceq \mathbb{I}$. In Subsection \ref{sec:quadratic-rounding}, we obtain such a matrix by solving a
semidefinite relaxation of~\eqref{eq: OPTProb}, using
the general framework for noncommutative polynomial
optimization developed in~\cite{PNA}. The relaxation's optimal value also provides
an upper bound on $\OPT(H)$, which we use to establish
approximation guarantees for the states we construct. 

To describe these relaxations, it is convenient to introduce Hermitian
\textit{Majorana operators}, defined for each $j\in V$ by 
\begin{equation}
\label{eq:majorana-definition}
    \chi_{2j-1}:=a_j+a_j^\dagger,
    \quad
    \chi_{2j}:=-i(a_j-a_j^\dagger).
\end{equation}
The CAR then imply
\begin{equation}
\label{eq:majorana-CAR}
    \chi_p^\dagger=\chi_p,
    \quad
    \chi_p\chi_q+\chi_q\chi_p=2\delta_{pq}\mathbb{I},
    \quad p,q\in\{1,\ldots,2n\}.
\end{equation}
Conversely, for $j\in V$,
$ a_j=\frac{\chi_{2j-1}+i\chi_{2j}}{2}$ and $a_j^\dagger=\frac{\chi_{2j-1}-i\chi_{2j}}{2}$. 
Since the original operators and the Majorana operators are
linear combinations of one another, either set can be used
to express all fermionic operators, with the same degree bounds.

Products of Majorana operators can be easily simplified using
$\chi_p^2=\mathbb{I}$ to eliminate repeated indices and
using $\chi_p\chi_q=-\chi_q\chi_p$ to arrange
the remaining ones  in increasing order.
For example, $\chi_3\chi_1\chi_3=-\chi_1\chi_3^2=-\chi_1$. 

Let $\mathbb C\langle\chi_1,\ldots,\chi_{2n}\rangle$ be the
free unital complex $*$-algebra generated by
$\chi_1,\ldots,\chi_{2n}$.
Let $\mathcal I_{\mathrm{Maj}}$ be the two-sided $*$-ideal
generated by the polynomials encoding the Majorana relations from \eqref{eq:majorana-CAR}.
The Majorana algebra is the complex Clifford algebra
(see, e.g.,~\cite[Chapter~I, Section~1]{LawsonMichelsohn1989}),
defined as the quotient
\begin{align}\nonumber
    \mathcal A_n
    :=\mathbb C\langle\chi_1,\ldots,\chi_{2n}\rangle
      /\mathcal I_{\mathrm{Maj}}.
\end{align}\noindent
For an ordered subset
$I=\{i_1<\cdots<i_m\}\subseteq\{1,\ldots,2n\}$, define
the Hermitian Majorana monomial
\begin{equation}\nonumber
    C_I:=i^{\binom{m}{2}}\chi_{i_1}\cdots\chi_{i_m},
    \quad C_\varnothing:=\mathbb{I},
\end{equation}
where the phase $i^{\binom{m}{2}}$ ensures that
$C_I^\dagger=C_I$ and $C_I^2=\mathbb{I}$.
The set
$\{C_I\;\mid\;I\subseteq\{1,\ldots,2n\}\}$ forms a basis of
$\mathcal A_n$, so every $F\in\mathcal A_n$ admits
a unique expansion
\begin{align}\nonumber
    F=\sum_{I\subseteq\{1,\ldots,2n\}}f_I C_I,
    \quad f_I\in\mathbb C.
\end{align}
For $d\ge0$, let $ \mathcal A_{\le d}$ be the subspace of Majorana monomials of degree at most $d$, i.e.,
\begin{equation}\nonumber
    \mathcal A_{\le d}
    :=\operatorname{span}_{\mathbb C}
      \{C_I\;\mid\;|I|\le d\}.
\end{equation}
Notice that, for example, $ n_j\in\mathcal A_{\le2},n_jn_k\in\mathcal A_{\le4}$ and $
    H\in\mathcal A_{\le4}$.
\begin{definition}[Truncated fermionic pseudo-moment functional]
\label{def:fermionic-moments}
For an integer $k\ge1$, let $\mathfrak L_k$ be the set of
complex linear functionals
$L:\mathcal A_{\le2k}\to\mathbb C$ satisfying
\begin{equation}
\label{eq:moment-conditions}
\begin{aligned}
    L(\mathbb{I})&=1,\\
    L(F^\dagger)&=\overline{L(F)},
        \quad\forall F\in\mathcal A_{\le2k},\\
    L(F^\dagger F)&\ge0,
        \quad\forall F\in\mathcal A_{\le k}.
\end{aligned}
\end{equation}
The values $\{L(C_I)\}_{|I|\le2k}$, are called 
the \emph{pseudo-moments} of $L$ of order $k$.
\end{definition}
\noindent
\begin{remark}\label{eq: pseudo-true}
The expectations $\langle C_I\rangle_\rho$ are called
the \emph{Majorana moments} of the state $\rho$. Every density matrix $\rho$ induces an element of
$\mathfrak L_k$ through
$L_\rho(F):=\Tr(\rho F)$ for $F\in\mathcal A_{\le2k}$.
However, a functional $L\in\mathfrak L_k$ does not
necessarily admit such a representation, since the truncated conditions~\eqref{eq:moment-conditions}
may not be sufficient to ensure that all the values
$L(C_I)$ arise from a single quantum state. 
We therefore call the values $L(C_I)$ \emph{pseudo-moments}
and use $L(F)$ to distinguish them from the state
expectations $\langle F\rangle_\rho$.
\end{remark}
We now show that the positivity condition in~\eqref{eq:moment-conditions}
has a finite matrix representation.
For $k\geq1$, let:
\begin{align}\nonumber
    \mathcal{B}_k:=\{C_I\;\mid\: |I|\leq k\}, \quad\text{with}\quad |\mathcal{B}_k|=\sum_{m=0}^{\min\{k,2n\}}\binom{2n}{m}=\mathcal{O}(n^k).
\end{align}
For $L\in\mathfrak L_k$, we define (entry-wise) the \textit{pseudo-moment (pseudo-correlation) matrix of order $k$} via 
\begin{align}\nonumber
    M^{(k)}_{I,J}(L):=L(C_I^\dagger C_j), \quad C_I,C_J\in\mathcal{B}_k.
    \end{align}
By construction, for any
$f=(f_I)_{|I|\leq k}$ and $F=\sum_{C_I\in\mathcal B_k}f_I C_I$, we have
\begin{equation}
\label{eq:moment-matrix-positivity}
\begin{aligned}\nonumber
    L(F^\dagger F)
    &=\sum_{|I|+|J|\leq 2k}\overline{f_I}f_J
        L(C_I^\dagger C_J)=f^\dagger M^{(k)}(L)f,
\end{aligned}
\end{equation}
and consequently,
\begin{align}\nonumber
   L(F^\dagger F)\ge0,\;  \forall F\in\mathcal A_{\le k}\;\Longleftrightarrow\;
    M^{(k)}(L)\succeq0.
\end{align}
We emphasize that the pseudo-moments $L(C_I)$ are real because the basis operators
$C_I$ are Hermitian. However, the products $C_I^\dagger C_J$
need not be Hermitian, so the pseudo-moment matrix may have complex
entries. For example, $L(\chi_1\chi_2)=-iL(C_{\{1,2\}})$. 

Since $H\in\mathcal A_{\le4}$,  $L(H)$ is
well defined whenever $k\ge2$.  We therefore define the \textit{fermionic moment relaxation of order $k\geq 2$} as follows:
\begin{equation}
\label{eq:SDP-definition}
    \operatorname{SDP}_k(H)
    :=\max_{L\in\mathfrak L_k}L(H).
\end{equation}
Since the conditions defining
$\mathfrak L_k$ consist of linear constraints and a
positive semidefinite matrix constraint, and since the objective is linear in the pseudo-moment variables $L(C_I)$, we conclude that the optimal value in  
\eqref{eq:SDP-definition} can be computed by solving a semidefinite program of size $|\mathcal B_k|$.
\begin{proposition}[Fermionic moment hierarchy of upper bounds]
\label{prop:moment-hierarchy}
For every Hamiltonian $H$ of the form~\eqref{eq: H-generic},
the maxima in~\eqref{eq:SDP-definition} are attained and satisfy
\begin{equation}
\label{eq:hierarchy-bounds}
    \operatorname{OPT}(H)
    =\operatorname{SDP}_{2n}(H)
    \le\cdots\le
    \operatorname{SDP}_3(H)
    \le\operatorname{SDP}_2(H).
\end{equation}
\end{proposition}
\begin{proof} From Remark \ref{eq: pseudo-true}, we deduce that $\operatorname{SDP}_k(H)
    \ge
    \operatorname{OPT}(H)$ for all $k\geq2$.
    
If $L\in\mathfrak L_{k+1}$, its restriction to
$\mathcal A_{\le2k}$ belongs to $\mathfrak L_k$ and preserves
the objective value. Increasing the relaxation level can therefore
only decrease the SDP optimum, establishing the monotonicity of the sequence $(\operatorname{SDP}_k(H))_{k\geq2}$.

Finally, at level $k=2n$, the space $\mathcal A_{\le 2n}$
coincides with the full operator algebra on $\mathcal F_n$,
so $L(F^\dagger F)\ge0$ for every operator $F$.
To see that $L$ comes from a quantum state, fix an
orthonormal basis $\{|\psi_r\rangle\}_{r=1}^{2^n}$ and define a matrix $\rho$ entry-wise via
$\rho_{rs}:=L(|\psi_s\rangle\langle \psi_r|)$.
Expanding any operator $F$ in this basis then gives
$L(F)=\operatorname{Tr}(\rho F)$. Furthermore, recall that every positive semidefinite operator $A$
admits a factorization $A=F^\dagger F$ for some
operator $F$. Feasibility then implies $L(A)\geq0$ for every $A\succeq0$. In particular,
$|\psi\rangle\langle \psi|\succeq0$ for every
$|\psi\rangle\in\mathcal F_n$, and hence
$\psi^\dagger\rho \psi
=\operatorname{Tr}(\rho|\psi\rangle\langle \psi|)
=L(|\psi\rangle\langle \psi|)\ge0$. 
Thus, $\rho\succeq0$ and $\Tr(\rho)=1$. Consequently, every feasible functional at this level is represented
by a density matrix, yielding $\operatorname{SDP}_{2n}(H)\leq\operatorname{OPT}(H)$. 

To prove attainment, note that, since $M^{(k)}(L)\succeq0$, all its $2\times2$
principal submatrices indexed by $I,J$ have nonnegative determinant. Thus, $|M^{(k)}_{I,J}(L)|\leq M^{(k)}_{I,I}(L)M^{(k)}_{J,J}(L)=L(C_I^2)L(C_J^2)=1$, i.e, the feasible set is closed and nonempty, hence compact.
The linear objective function consequently attains its maximum.
\end{proof}
Approximation guarantees that will be considered in the rest of this paper exploit the minimal relaxation level $k=2$.  
At this level, $|\mathcal B_2|
    =1+2n+\binom{2n}{2}=\mathcal{O}(n^2)$. 
Therefore, if an operator $P\in\mathcal A_{\le4}$ admits a decomposition $P=\sum_{|I|\leq 2} F_I^\dagger F_I,
    \; F_I\in\mathcal A_{\le2}$, 
then $L(P)\ge0$ for all feasible $L\in \mathfrak L_2$. 

In particular, let $L_\star\in\mathfrak L_2$ be an optimal solution of \eqref{eq:SDP-definition} for $k=2$.
Then,
\begin{equation}
\label{eq:optimal-moment-functional}
    L_\star(H)
    =\operatorname{SDP}_2(H)
    \ge\operatorname{OPT}(H).
\end{equation}
In the next section, we show how to turn the pseudo-moments
of an optimal solution $L_\star$ into Gaussian states whose
energies retain a constant fraction of the relaxation value $L_\star(H)$.

\section{Gaussian states from pseudo-moments}
\label{sec:rounding}
We now introduce two rounding schemes that use different
pseudo-moments of the same solution $L_\star$ to construct
valid Gaussian states.
The first one, described in Subsection  \ref{sec:quadratic-rounding}, preserves the occupation and hopping values
specified by $L_\star$, but Gaussianity forces the interaction expectations to follow Wick's formula.
This can reduce the interaction energy compared with
the SDP objective, motivating a second scheme, detailed in Subsection \ref{sec:occupation-rounding}, that directly relies on
the joint occupation pseudo-moments.
It constructs occupation-basis states, whose hopping
expectations vanish. We further analyze both schemes to be able to control how much of the relaxation value their expected energies
retain.

Recall that $\langle F\rangle_\rho=\operatorname{Tr}(\rho F)$
denotes the quantum expectation of $F$ in the state $\rho$.
For a pure state $\rho=|\psi\rangle\langle\psi|$, we use
$\langle F\rangle_\psi=\langle\psi|F|\psi\rangle$
interchangeably with $\langle F\rangle_\rho$.
\subsection{Rounding one-particle correlations}
\label{sec:quadratic-rounding}
The quadratic contribution to the SDP objective $L_\star(H)$
depends only on the pseudo-moments of the form $L_\star(a_j^\dagger a_k)$ with $ j,k\in V$.
We therefore define
\begin{equation}
\label{eq:SDP-R}
    R_{jk}:=L_\star(a_j^\dagger a_k),
    \quad j,k\in V.
\end{equation}
If $L_\star$ came from a density matrix, $R$ would be its
one-particle correlation matrix. Since $L_\star$ does not necessarily come from a quantum state, we must verify that $R$ still
satisfies the hypotheses needed to invoke
Proposition~\ref{prop:gaussian-realization}.
The next proposition establishes this.
\begin{proposition}[SDP-based one-particle matrix]
\label{prop:SDP-R}
The matrix $R$ in~\eqref{eq:SDP-R} satisfies
\begin{align}\nonumber
R=R^\dagger,
    \qquad 0\preceq R\preceq \mathbb{I}.
\end{align}
\end{proposition}
\begin{proof}
Since $L_\star$ satisfies $L_\star(F^\dagger)=\overline{L_\star(F)}$ for $F\in\mathcal{A}_{\leq 2}$,
we have $R_{kj}
=L_\star(a_k^\dagger a_j)
=\overline{L_\star(a_j^\dagger a_k)}
=\overline{R_{jk}}$, i.e., $R$ is Hermitian.

For $z\in\mathbb C^n$, set $b:=\sum_jz_ja_j$.
Since $b,b^\dagger\in\mathcal{A}_{\leq 1}$,  the positivity condition 
in~\eqref{eq:moment-conditions} gives $z^\dagger Rz=L_\star(b^\dagger b)\ge0$. 
The CAR further imply
$bb^\dagger+b^\dagger b=\|z\|_2^2\mathbb{I}$.
Using $L_\star(\mathbb{I})=1$, we consequently obtain
$z^\dagger(\mathbb{I}-R)z=L_\star(bb^\dagger)\ge0$, which ends the proof.
\end{proof}
Invoking Proposition~\ref{prop:gaussian-realization}, we deduce that there exists a number-conserving Gaussian state $\rho_R$ satisfying
\begin{align}\nonumber
\langle a_j^\dagger a_k\rangle_{\rho_R}
    =R_{jk}=L_\star(a_j^\dagger a_k).
\end{align}
Let us write  $R=U\operatorname{diag}(\nu_1,\ldots,\nu_n)U^\dagger$ as in \eqref{eq:R-eigendecomposition},  and  define $b_\ell,b_\ell^\dagger$ as in
\eqref{eq:normal-modes}. Our first rounding scheme thus reduces to independently sampling
the occupations $x_\ell\in\{0,1\}$  with
$\Pr(x_\ell=1)=\nu_\ell$ for each $\ell\in V$,
and returning the following state
\begin{equation}
\label{eq:SDP-Q-state}
    |\psi_{\mathrm{SDP-Q}}\rangle
    :=(b_1^\dagger)^{x_1}\cdots
      (b_n^\dagger)^{x_n}|\mathrm{vac}\rangle.
\end{equation}
Each sampling outcome occupies a collection of orthonormal modes. By  Remark~\ref{rem:Slater}, it therefore represents a Slater determinant, hence a pure
Gaussian state. Moreover, the probability of sampling a particular occupation pattern
$x\in\{0,1\}^n$ is exactly $p_R(x)$ from
\eqref{eq:gaussian-realization}, and consequently,
\begin{equation}
\label{eq:SDP-Q-average}
    \mathbb E\!\left[
        |\psi_{\mathrm{SDP-Q}}\rangle
        \langle\psi_{\mathrm{SDP-Q}}|
    \right]
=\sum_{x\in\{0,1\}^n}p_R(x)|\psi_x\rangle\langle\psi_x|
=\rho_R.
\end{equation} 
\begin{proposition}
\label{prop:SDP-Q-moments}
The Gaussian state in~\eqref{eq:SDP-Q-state} satisfies:
\begin{enumerate}
    \item[(i)] For all $j,k\in V$,
    \begin{equation}
    \label{eq:SDP-Q-quadratic}
        \mathbb E\!\left[
            \langle a_j^\dagger a_k
            \rangle_{\psi_{\mathrm{SDP-Q}}}
        \right]
        =R_{jk}=L_\star(a_j^\dagger a_k).
    \end{equation}
    \item[(ii)] For distinct modes $j,k\in V$,
    \begin{equation}
    \label{eq:SDP-Q-quartic}
        \mathbb E\!\left[
            \langle n_jn_k\rangle_{\psi_{\mathrm{SDP-Q}}}
        \right]
        =R_{jj}R_{kk}-|R_{jk}|^2.
    \end{equation}
    \item[(iii)] Its expected energy $E_{\mathrm Q}
        :=\mathbb E\!\left[
            \langle H\rangle_{\psi_{\mathrm{SDP-Q}}}\right]$ is
    \begin{equation}
    \label{eq:SDP-Q-energy}
    \begin{aligned}
        E_{\mathrm Q}=\gamma+\sum_{j\in V}\mu_jR_{jj}
          -\sum_{\{j,k\}\in E}w_{jk}\bigl(R_{jj}R_{kk}-|R_{jk}|^2\bigr)
          -2\sum_{\{j,k\}\in E'}w'_{jk}\operatorname{Re}(R_{jk}).
    \end{aligned}
    \end{equation}
\end{enumerate}
\end{proposition}
\begin{proof}
Linearity of $\Tr$ and \eqref{eq:SDP-Q-average} imply that $\mathbb E\!\left[
        \langle F\rangle_{\psi_{\mathrm{SDP-Q}}}
    \right]
    =\Tr(\E\left[
        |\psi_{\mathrm{SDP-Q}}\rangle
        \langle\psi_{\mathrm{SDP-Q}}|
    \right]F)=\operatorname{Tr}(\rho_R F)$
for every operator $F\in\mathcal{F}_n$.  
Taking $F=a_j^\dagger a_k$ gives
\eqref{eq:SDP-Q-quadratic} by
Proposition~\ref{prop:gaussian-realization}.
Furthermore, taking $F=n_jn_k$ gives
\eqref{eq:SDP-Q-quartic} by
Proposition~\ref{prop:wick-occupations}.
Finally, substitution into~\eqref{eq: H-generic} proves
\eqref{eq:SDP-Q-energy}.
\end{proof}
\noindent
All quadratic Hamiltonian terms contribute equally to
$E_Q$ and $L_\star(H)$. Only the interaction terms
remain when we subtract the two, which gives
\begin{equation}
\label{eq:SDP-Q-energy-difference}
    E_{\mathrm Q}
    =L_\star(H)
     +\sum_{\{j,k\}\in E}w_{jk}
       \left(
           L_\star(n_jn_k)-R_{jj}R_{kk}+|R_{jk}|^2
       \right).
\end{equation}
Any energy loss relative to $L_\star(H)$ therefore comes
solely from replacing $L_\star(n_jn_k)$ by the Wick values
$R_{jj}R_{kk}-|R_{jk}|^2$, and our subsequent approximation
analysis focuses on bounding this loss precisely.

We finally emphasize that a sampled state $|\psi_{\rm SDP-Q}\rangle$ has a definite particle number
$\sum_{\ell=1}^n x_\ell$, which may vary between samples. However, the expected particle number coincides
with the value assigned to the total particle-number
operator $N$ by the SDP solution:
\begin{equation}
\label{eq:SDP-Q-particle-number}
    \mathbb E\!\left[
        \langle N\rangle_{\psi_{\mathrm{SDP-Q}}}
    \right]
    =\sum_{\ell=1}^n\E[\langle n_{\ell}\rangle_{\psi_{\mathrm{SDP-Q}}}]=\sum_{\ell=1}^n\mathbb E[x_\ell]
    =\sum_{\ell=1}^n\nu_\ell=\operatorname{Tr}(R)
    =L_\star(N).
\end{equation}

\subsection{Rounding occupation correlations}
\label{sec:occupation-rounding}
The first scheme relies only on quadratic terms and does not take
the joint occupation pseudo-moments $L_\star(n_jn_k)$
into account. In contrast, our second scheme incorporates this information to
construct a random
occupation pattern through standard hyperplane rounding \cite{GW}.

For each mode $j\in V$, let
$Z_j:=\mathbb{I}-2n_j.$
The value of $Z_j$ on an occupation-basis state is $1$
for an empty mode and $-1$ for an occupied mode.
Thus, for $k\neq j$, the quartic operator $Z_jZ_k$ equals $1$ when the occupations agree
and $-1$ when they differ.
This motivates the introduction of the following \textit{occupation-agreement matrix}
\begin{equation}
\label{eq:S-R-quartic}
S_{jk}:=L_\star(Z_jZ_k)=1-2R_{jj}-2R_{kk}+4L_\star(n_jn_k),
    \quad j,k\in V.
\end{equation}
Unlike $R$, the matrix $S$ therefore includes both joint and individual occupation pseudo-moments.

\begin{proposition}[SDP-based occupation agreement matrix]
\label{prop:SDP-S}
The matrix $S$ from \eqref{eq:S-R-quartic} is real symmetric, positive semidefinite,
and has unit diagonal. Consequently, there exist 
unit vectors $v_1,\ldots,v_n\in\mathbb R^n$ such that $S_{jk}=v_j^\top v_k$ for all $j,k\in V$.
\end{proposition}
\begin{proof}
Since the occupation operators commute and satisfy
$n_j^2=n_j$, we have the following relations:
\begin{align}\nonumber
     Z_jZ_k=Z_kZ_j,
    \quad
    (Z_jZ_k)^\dagger=Z_jZ_k,
    \quad Z_j^2=\mathbb{I}.
\end{align}
Combining these identities with the defining properties
of $L_\star$ shows that $S$ is real symmetric and satisfies
$S_{jj}=L_\star(\mathbb{I})=1$ for every $j\in V$.

For $z\in\mathbb R^n$, define $F_z:=\sum_{j\in V} z_jZ_j\in\mathcal{A}_{\leq 2}$. The level-two
positivity implies $z^\top Sz=L_\star(F_z^\dagger F_z)\ge0$, and hence $S\succeq0$.
A real Gram factorization applied to $S$ then gives vectors $\{v_j\}_{j\in V}$
that satisfy $v_j^\top v_k=S_{jk}$, and  $||v_j||=S_{jj}=1$ for all $j\in V$.
\end{proof}
The Gram factorization from Proposition \ref{prop:SDP-S} enables us to apply the random hyperplane
rounding procedure from \cite{GW}. Namely, we first sample a vector $g$ from the normal distribution $\mathcal N(0,\mathbb{I})$ and then let
\begin{equation}\nonumber
    y_j:=\frac{1-\operatorname{sign}(g^\top v_j)}{2}\in\{0,1\},
    \quad j\in V.
\end{equation}
We can then 
construct the following quantum state:
\begin{equation}
\label{eq:SDP-cut-state}
    |\psi_{\mathrm{SDP-cut}}\rangle
    :=|y_1\cdots y_n\rangle=(a_1^\dagger)^{y_1}\cdots
      (a_n^\dagger)^{y_n}|\mathrm{vac}\rangle.
\end{equation}
By Remark~\ref{rem:Slater}, each realization of $|\psi_{\rm SDP-cut}\rangle$ is 
a Slater determinant.

We next compute the expected occupation and hopping moments
of $|\psi_{\mathrm{SDP-cut}}\rangle$ by exploiting the key identity $\Pr(y_j\ne y_k)=\frac{\arccos(S_{jk})}{\pi}$ \cite{GW}, 
which gives the probability that the sampled hyperplane, defined by $g$,
separates two distinct Gram vectors $v_j$ and $v_k$.

\begin{proposition}
\label{prop:SDP-cut-moments}
The Gaussian state in~\eqref{eq:SDP-cut-state} satisfies:
\begin{enumerate}
    \item[(i)] For every $j\in V$,
    \begin{equation}
    \label{eq:SDP-cut-quadratic}
        \mathbb E\!\left[
            \langle n_j\rangle_{\psi_{\mathrm{SDP-cut}}}
        \right]=\frac12,
        \quad
        \langle a_j^\dagger a_k
        \rangle_{\psi_{\mathrm{SDP-cut}}}=0, 
        \; \forall k\ne j.
    \end{equation}

    \item[(ii)] For distinct modes $j,k\in V$,
    \begin{align}
    \label{eq:SDP-cut-disagreement}
        \mathbb E\!\left[
            \langle n_jn_k\rangle_{\psi_{\mathrm{SDP-cut}}}
        \right]
        =\frac12
          \left(1-\frac{\arccos(S_{jk})}{\pi}\right).
    \end{align}

    \item[(iii)] Its expected energy $E_{\mathrm{cut}}
        :=\mathbb E\!\left[
            \langle H\rangle_{\psi_{\mathrm{SDP-cut}}}
        \right]$ is
    \begin{equation}
    \label{eq:SDP-cut-energy}
    \begin{aligned}
        E_{\mathrm{cut}}=\gamma+\frac12\sum_{j\in V}\mu_j
          -\frac12\sum_{\{j,k\}\in E}w_{jk}
            \left(1-\frac{\arccos(S_{jk})}{\pi}\right).
    \end{aligned}
    \end{equation}
\end{enumerate}
\end{proposition}
\begin{proof} \textit{(i)} Since $\|v_j\|=1$, the scalar $g^\top v_j$ has the standard
normal distribution for each $j\in V$. Therefore, $\mathbb E\!\left[
        \langle n_j\rangle_{\psi_{\mathrm{SDP-cut}}}
    \right]
    =\Pr(y_j=1)=\frac12$. For $j\ne k$, diagonal entries of $a_j^\dagger a_k$ 
in the occupation-number basis are all equal to zero. Hence, 
$\langle a_j^\dagger a_k\rangle_{\psi_{\mathrm{SDP-cut}}}=0$
for every realization of $|\psi_{{\mathrm{SDP-cut}}}\rangle$.

\textit{(ii)} By the analysis of hyperplane rounding
in~\cite{GW}, $\Pr(y_j\ne y_k)
    =\frac{\arccos(v_j^\top v_k)}{\pi}
    =\frac{\arccos(S_{jk})}{\pi}$. Moreover, for binary variables,
$\mathbf{1}_{\{y_j\ne y_k\}}=y_j+y_k-2y_jy_k$.
Taking expectations and using part~\textit{(i)} therefore gives
$ \mathbb E\!\left[
        \langle n_jn_k\rangle_{\psi_{\mathrm{SDP-cut}}}
    \right]
    =\mathbb E[y_jy_k]
    =\frac12\left(
        1-\frac{\arccos(S_{jk})}{\pi}
      \right)$.
      
\textit{(iii)} Straightforward, by
substituting the preceding expressions into
\eqref{eq: H-generic}.
\end{proof}

\subsection{Combining the rounding schemes}
\label{sec:rounding-selection}
To approximate the optimal value of the unconstrained problem~\eqref{eq: OPTProb}, we sample
one state from each of the two preceding rounding procedures and
return the one with larger energy. Thus,  our approximation algorithm outputs
\begin{equation}
\label{eq:state-selection}
|\psi_{\mathrm{alg}}\rangle:=
    \begin{cases}
        |\psi_{\mathrm{SDP-Q}}\rangle,
        & \text{if}\quad\langle H\rangle_{\psi_{\mathrm{SDP-Q}}}
          \ge \langle H\rangle_{\psi_{\mathrm{SDP-cut}}},\\
        |\psi_{\mathrm{SDP-cut}}\rangle,
        & \text{otherwise}.
    \end{cases}
\end{equation}
Both candidates are Slater determinants, so every output
is a pure number-conserving Gaussian state. Jensen's inequality applied to the convex function $(x,y)\mapsto\max\{x,y\}$, together with the fact that a convex combination of two numbers cannot exceed
their maximum, yields the following inequality:
\begin{equation}
\label{eq:best-candidate-bound}
    \mathbb E\!\left[
        \langle H\rangle_{\psi_{\mathrm{alg}}}
    \right]
    \ge \max\{E_{\mathrm Q},E_{\mathrm{cut}}\}
    \ge \lambda E_{\mathrm Q}
       +(1-\lambda)E_{\mathrm{cut}},
    \quad \lambda\in[0,1].
\end{equation}
Hence, to obtain a constant approximation guarantee for a given Hamiltonian family $\mathcal{H}$, it suffices to
find $\lambda\in[0,1]$ and $\alpha\in(0,1]$ such that for all $H\in\mathcal H$,
\begin{equation}
\label{eq:rounding-SDP-guarantee}
    \lambda E_{\mathrm Q}
    +(1-\lambda)E_{\mathrm{cut}}
    \ge \alpha L_\star(H).
\end{equation}
Indeed, together with~\eqref{eq:optimal-moment-functional}, this would then imply the $\alpha$-approximation ratio, i.e., 
\begin{align}\nonumber
    \mathbb E\!\left[
        \langle H\rangle_{\psi_{\mathrm{alg}}}
    \right]
    \ge \alpha L_\star(H)
    \ge \alpha\operatorname{OPT}(H).
\end{align}
\noindent

\begin{remark}[On computation]\label{rem: computation}
The one-particle correlation matrices of our two sampled
states are $U\operatorname{diag}(x_1,\ldots,x_n)U^\dagger$
and $\operatorname{diag}(y_1,\ldots,y_n)$, respectively.
Their energies can therefore be evaluated
using~\eqref{eq:gaussian-energy}, without forming full
density matrices or performing quantum measurements.

Since the relaxation order is fixed at $k=2$,
the corresponding SDP has $\mathcal O(n^4)$ scalar
decision variables subject to a linear matrix
    inequality of size $\mathcal O(n^2)$. Given a solution of this SDP,
the subsequent rounding procedures require only
the spectral decomposition of $R$, the Gram
factorization of $S$, sampling, and energy evaluation,
all involving matrices of dimension at most $n$.

Finally, our approximation guarantees are established for
optimal SDP solutions. The effects of finite-precision
computation on these guarantees are not analyzed here.
In this regard, the numerical certification of SDP
solutions arising in quantum optimization has recently
been studied in~\cite{Naceur2026}.
\end{remark}
%%%%%%%%%%%%%%%%%%%%%%%%%%%%%%%%%%%%%%%%%%%%%%%%%%%%%%%%
\subsection{Properties of two-mode pseudo-moments}
To compare the energies of the two rounding schemes
with the SDP objective, we first establish constraints
on the pseudo-moments associated with a pair of modes.
These results follow from level-two positivity and do not depend
on the Hamiltonian's coefficients or normalization.
\begin{lemma}
\label{lem:PSD-occupation-moments}
Let $L\in\mathfrak L_2$. For distinct modes $j,k\in V$, write
\[
r_j:=L(n_j),\quad
r_k:=L(n_k),\quad
d:=L(n_jn_k),\quad
s:=L(Z_jZ_k),\quad\text{and}\quad
z:=L(a_j^\dagger a_k).
\]
Then, the following inequalities hold:
\begin{equation}
\label{eq:PSD-occupation-bounds}
    d\ge0,\quad
    r_j-d\ge0,\quad
    r_k-d\ge0,\quad\text{and}\quad
    1-r_j-r_k+d\ge0.
\end{equation}
Moreover,
\begin{equation}
\label{eq:PSD-joint-occupation-loss}
    r_jr_k-|z|^2-d
    \le r_jr_k-d
    \le\frac{(1-s)^2}{16}.
\end{equation}
In particular, when $L=L_\star$, both \eqref{eq:PSD-occupation-bounds} and \eqref{eq:PSD-joint-occupation-loss} hold for
$(r_j,r_k,d,s,z)
=(R_{jj},R_{kk},L_\star(n_jn_k),S_{jk},R_{jk})$.
\end{lemma}

\begin{proof}
The four inequalities in~\eqref{eq:PSD-occupation-bounds}
follow by applying level-two positivity of $L$ to the following squares:
\[
\begin{aligned}
    n_jn_k&=(a_ka_j)^\dagger(a_ka_j),&
    n_j(\mathbb{I}-n_k)&=(a_k^\dagger a_j)^\dagger(a_k^\dagger a_j),\\
    (\mathbb{I}-n_j)n_k&=(a_j^\dagger a_k)^\dagger(a_j^\dagger a_k),&
    (\mathbb{I}-n_j)(\mathbb{I}-n_k)&=(a_j^\dagger a_k^\dagger)^\dagger
                         (a_j^\dagger a_k^\dagger).
\end{aligned}
\]
The first inequality in~\eqref{eq:PSD-joint-occupation-loss}
follows from $|z|^2\ge0$.
For the second, expanding $Z_jZ_k=\frac{\mathbb{I}-n_j}{2}\frac{\mathbb{I}-n_k}{2}$ and applying $L$ gives
$s=1-2r_j-2r_k+4d$. Hence,
\[
\begin{aligned}
    \frac{(1-s)^2}{16}-r_jr_k+d
    &=\left(\frac{r_j+r_k}{2}-d\right)^2-r_jr_k+d=\frac{(r_j-r_k)^2}{4}
      +d(1-r_j-r_k+d).
\end{aligned}
\]
Both factors in the last product
being nonnegative by~\eqref{eq:PSD-occupation-bounds}, we deduce that the difference is nonnegative as well, proving the desired result.
\end{proof}
Bounding the rounding loss on each interaction
edge requires knowing which combinations of occupation and
hopping pseudo-moments are jointly feasible. The next lemma
describes this feasible set exactly.
\begin{lemma}
\label{lem:two-mode-domain}
For distinct $j,k\in V$, define
\begin{equation}
\label{eq:two-mode-domain}
\mathcal D_{jk}:=
\left\{
(r_j,r_k,d,s,z)\in\mathbb R^4\times\mathbb C
\;\middle|\;
\begin{aligned}
d&\ge0,\\
1-r_j-r_k+d&\ge0,\\
\begin{pmatrix}
r_j-d & \overline z\\
z & r_k-d
\end{pmatrix}&\succeq0,\\
1-2r_j-2r_k+4d&=s
\end{aligned}
\right\}.
\end{equation}
Then, for every $L\in\mathfrak L_2$, we have $\bigl(
L(n_j),L(n_k),L(n_jn_k),
L(Z_jZ_k),L(a_j^\dagger a_k)
\bigr)\in\mathcal D_{jk}$. Conversely, for every point $(r_j,r_k,d,s,z)\in\mathcal D_{jk}$ there exists a two-mode density matrix $\rho_{jk}\in\mathbb{C}^{4\times4}$ satisfying
\[
\bigl(
\langle n_j\rangle_{\rho_{jk}},
\langle n_k\rangle_{\rho_{jk}},
\langle n_jn_k\rangle_{\rho_{jk}},
\langle Z_jZ_k\rangle_{\rho_{jk}},
\langle a_j^\dagger a_k\rangle_{\rho_{jk}}
\bigr)
=(r_j,r_k,d,s,z).
\]
\end{lemma}

\begin{proof}
Fix $L\in\mathfrak L_2$ and write
\[
r_j:=L(n_j),\quad
r_k:=L(n_k),\quad
d:=L(n_jn_k),\quad
s:=L(Z_jZ_k),\quad\text{and}\quad
z:=L(a_j^\dagger a_k).
\]
The first four quantities above are real, since
$n_j$, $n_k$, $n_jn_k$, and $Z_jZ_k$ are Hermitian. Next, the constraints $d\geq0$ and $1-r_j-r_k+d\ge0$ follow from
level-two positivity as in
Lemma~\ref{lem:PSD-occupation-moments}, while expanding
$Z_jZ_k=(\mathbb{I}-2n_j)(\mathbb{I}-2n_k)$ gives
$s=1-2r_j-2r_k+4d$.

It remains to prove the $2\times 2$ linear matrix inequality
involving the hopping quantity $z$. Let $u,v\in\mathbb C$
satisfy $|u|^2+|v|^2=1$ and define
$b_1:=\overline u\,a_j+\overline v\,a_k$ and
$b_2:=-v\,a_j+u\,a_k$.
These operators satisfy the CAR, and $b_2b_1=a_ka_j$.
Consequently, $(b_2^\dagger b_1)^\dagger(b_2^\dagger b_1)
=b_1^\dagger b_1-n_jn_k$, with
$b_1^\dagger b_1
    =|u|^2n_j+|v|^2n_k
     +u\overline v\,a_j^\dagger a_k
     +v\overline u\,a_k^\dagger a_j$.
Since $b_2^\dagger b_1\in\mathcal A_{\le2}$,
level-two positivity gives
\begin{align*}
0
\le L\!\left(
(b_2^\dagger b_1)^\dagger(b_2^\dagger b_1)
\right)=|u|^2r_j+|v|^2r_k
  +u\overline v\,z+v\overline u\,\overline z-d =
\begin{pmatrix}u\\v\end{pmatrix}^{\!\dagger}
\begin{pmatrix}
r_j-d&\overline z\\
z&r_k-d
\end{pmatrix}
\begin{pmatrix}u\\v\end{pmatrix}.
\end{align*}
This holds for every unit vector $(u,v)$, so the matrix
in the last line is positive semidefinite.

Conversely, fix any point $(r_j,r_k,d,s,z)\in\mathcal D_{jk}$, and define the following Hermitian matrix in the occupation basis
$|00\rangle,|10\rangle,|01\rangle,|11\rangle$:
\[
\rho_{jk}:=
\begin{pmatrix}
1-r_j-r_k+d&0&0&0\\
0&r_j-d&\overline z&0\\
0&z&r_k-d&0\\
0&0&0&d
\end{pmatrix}.
\]
Since $(r_j,r_k,d,s,z)\in\mathcal D_{jk}$, we deduce that  $\rho_{jk}\succeq0$. Moreover, $\Tr(\rho_{jk})=1$, so $\rho_{jk}$ is a density matrix. Direct computation gives
\[
\langle n_j\rangle_{\rho_{jk}}=r_j,\quad
\langle n_k\rangle_{\rho_{jk}}=r_k,\quad
\langle n_jn_k\rangle_{\rho_{jk}}=d,\quad
\langle a_j^\dagger a_k\rangle_{\rho_{jk}}=z.
\]
Finally, $\langle Z_jZ_k\rangle_{\rho_{jk}}
=1-2r_j-2r_k+4d=s$, 
which completes the proof.
\end{proof}

\section{Approximation guarantees for positive semidefinite Hamiltonians}
\label{sec:PSD-guarantees}
In this section, we establish constant approximation guarantees for two
positive semidefinite models, without any assumptions on the
interaction graph $(V,E)$. Following the normalization from \cite{STH}, we first consider Hamiltonians whose
occupation, interaction, and hopping terms are individually
positive semidefinite. We then adapt the analysis to Fermionic Max Cut.

\subsection{The general termwise positive semidefinite model}
\label{sec:PSD-model}

Consider
\begin{equation}
\label{eq:H-PSD}
\begin{aligned}
    H_{\mathrm{PSD}}
    :={}&\sum_{\{j,k\}\in E'}w'_{jk}
      \bigl(\mathbb{I}-a_j^\dagger a_k-a_k^\dagger a_j\bigr)+
    \sum_{\{j,k\}\in E}w_{jk}(\mathbb{I}-n_jn_k)
    +\sum_{j\in V}\mu_jn_j,
\end{aligned}
\end{equation}
where we suppose $w_{jk}',w_{jk},\mu_j\ge0$.
Thus, $H_{\rm PSD}$ is of the form \eqref{eq: H-generic}, with $ \gamma
    =\sum_{\{j,k\}\in E}w_{jk}
     +\sum_{\{j,k\}\in E'}w'_{jk}$.  Since $n_j$ and $n_jn_k$ are orthogonal projectors,
both $n_j$ and $\mathbb{I}-n_jn_k$ are positive semidefinite.
Moreover, the shifted hopping terms are also positive semidefinite,
which follows from the CAR identity
\begin{equation}
\label{eq:PSD-hopping-squares}
\begin{aligned}
    \mathbb{I}\pm(a_j^\dagger a_k+a_k^\dagger a_j)
    =\frac12\Bigl[
       &(a_j\pm a_k)^\dagger(a_j\pm a_k) +(a_j^\dagger\mp a_k^\dagger)^\dagger
         (a_j^\dagger\mp a_k^\dagger)
    \Bigr]\succeq 0.
\end{aligned}
\end{equation}
Thus,  $H_{\mathrm{PSD}}\succeq0$, since every summand in~\eqref{eq:H-PSD} is positive
semidefinite.

Throughout this subsection, $R$, $S$, $E_{\mathrm Q}$,
and $E_{\mathrm{cut}}$ are constructed as in
Section~\ref{sec:rounding} from an optimal solution $L_\star\in\mathfrak L_2$
of the level-two relaxation for $H_{\mathrm{PSD}}$.

Using Proposition~\ref{prop:SDP-Q-moments} and Proposition~\ref{prop:SDP-cut-moments}, we begin by summarizing the expected energy contributions of each Hamiltonian term $T_{jk}\in\{n_j,\ \mathbb{I}-a_j^\dagger a_k-a_k^\dagger a_j,\ \mathbb{I}-n_jn_k\}$
from \eqref{eq:H-PSD}:
\begin{center}
\renewcommand{\arraystretch}{1.6}
\begin{tabular}{@{}llll@{}}
\toprule
$T_{jk}$ & $L_\star(T_{jk})$
  & $\E\left[\langle T_{jk}\rangle_{\psi_{\rm SDP-Q}}\right]$ & $\E\left[\langle T_{jk}\rangle_{\psi_{\rm SDP-cut}}\right]$\\
\midrule
$n_j$ & $R_{jj}$ & $R_{jj}$ & $\frac12$\\
$\mathbb{I}-a_j^\dagger a_k-a_k^\dagger a_j$
  & $1-2\operatorname{Re}R_{jk}$
  & $1-2\operatorname{Re}R_{jk}$ & $1$\\
$\mathbb{I}-n_jn_k$ & $1-L_\star(n_jn_k)$
  & $1-R_{jj}R_{kk}+|R_{jk}|^2$
  & $\frac12\left(1+\frac{\arccos(S_{jk})}{\pi}\right)$\\
\bottomrule
\end{tabular}
\end{center}
Following \eqref{eq:rounding-SDP-guarantee}, an $\alpha$-approximation ratio would be obtained if there existed $\lambda\in[0,1]$ such that
\begin{equation}
\label{eq:PSD-termwise-comparison}
\begin{aligned}
   \lambda R_{jj}+\frac{1-\lambda}{2}
        &\ge\alpha R_{jj},
       \quad \forall j\in V,\\
    \lambda(1-2\operatorname{Re}R_{jk})+(1-\lambda)
        &\ge\alpha(1-2\operatorname{Re}R_{jk}),
       \quad\forall\{j,k\}\in E',\\
    \lambda(1-R_{jj}R_{kk}+|R_{jk}|^2)
       +\frac{1-\lambda}{2}
          \left(1+\frac{\arccos(S_{jk})}{\pi}\right) & \ge\alpha\bigl(1-L_\star(n_jn_k)\bigr),
       \quad\forall\{j,k\}\in E.
\end{aligned}
\end{equation}
Indeed, multiplying these inequalities by nonnegative weights $\mu_j$, $w'_{jk}$, and
$w_{jk}$, respectively, and summing yields $\lambda E_{\mathrm Q}+(1-\lambda)E_{\mathrm{cut}}
    \ge\alpha L_\star(H_{\mathrm{PSD}})$. 
The approximation guarantee then follows from equations
\eqref{eq:best-candidate-bound} and \eqref{eq:optimal-moment-functional}.

For the first two rows, the comparison is simple, since
\begin{align}
\label{eq:PSD-occupation-ratio}
    \lambda R_{jj}+\frac{1-\lambda}{2}
    &\ge\left(\lambda+\frac{1-\lambda}{2}\right)R_{jj},\\
\label{eq:PSD-hopping-ratio}
    \lambda(1-2\operatorname{Re}R_{jk})+(1-\lambda)
    &\ge\left(\lambda+\frac{1-\lambda}{2}\right)
       (1-2\operatorname{Re}R_{jk}),
\end{align}
where the first inequality uses $R_{jj}\le1$, and the second one exploits 
$0\le1-2\operatorname{Re}R_{jk}\le2$, which follows from
level-two positivity and~\eqref{eq:PSD-hopping-squares}.
Thus, these two types of Hamiltonian terms each retain the fraction
$\lambda+\frac{1-\lambda}{2}=\frac{1+\lambda}{2}$ of their associated SDP values, a quantity which increases with $\lambda$. 

The interaction terms, on the other hand, impose a competing
constraint on the choice of $\lambda$. By
Lemma~\ref{lem:PSD-occupation-moments}, replacing
$L_\star(n_jn_k)$ by $R_{jj}R_{kk}-|R_{jk}|^2$ may reduce
the contribution of $\mathbb{I}-n_jn_k$ by at most
$\frac{(1-S_{jk})^2}{16}$. We use the following scalar inequality to compare this possible
loss with the contribution
$\frac{\arccos(S_{jk})}{2\pi}$ from hyperplane rounding
and choose $\lambda$ accordingly.
\begin{lemma}
\label{lem:PSD-scalar-comparison}
Let
\begin{equation}
\label{eq:PSD-kappa}
    \kappa:=\inf_{0<t\le1}
       \frac{\arccos(1-2t)}{\pi t^2}\approx0.94501.
\end{equation}
Then, for every $s\in[-1,1]$,
\begin{equation}\label{eq:PSD-scalar-comparison}
    \frac{\arccos(s)}{\pi}\ge\frac{\kappa}{4}(1-s)^2.
\end{equation}
\end{lemma}
\begin{proof}
At $s=1$, both
sides of~\eqref{eq:PSD-scalar-comparison} are zero. For $-1\leq s<1$, set $t=\frac{1-s}{2}\in]0,1]$. The definition of $\kappa$
gives
\[
    \kappa\le\frac{\arccos(1-2t)}{\pi t^2}
      =\frac{4\arccos(s)}{\pi(1-s)^2}.
\]
Multiplying by $\frac{(1-s)^2}{4}$ then proves the claim.
\end{proof}
For a fixed weight $\lambda\in[0,1]$,  let us define the worst-case interaction ratio by
\begin{equation}
\label{eq:PSD-interaction-optimization}
\beta(\lambda):=
\inf_{\substack{
    (r_j,r_k,d,s,z)\in\mathcal D_{jk},\\
    d<1
}}
\frac{
    \lambda\bigl(1-r_jr_k+|z|^2\bigr)
    +\dfrac{1-\lambda}{2}
       \left(1+\dfrac{\arccos(s)}{\pi}\right)
}{1-d}.
\end{equation}
In other words, for a fixed $\lambda$, the scalar $\beta(\lambda)$ represents the largest constant
such that, for every $(r_j,r_k,d,s,z)\in\mathcal D_{jk}$ with $d<1$,
\[
\lambda(1-r_jr_k+|z|^2)
+\frac{1-\lambda}{2}
 \left(1+\frac{\arccos(s)}{\pi}\right)
\ge \beta(\lambda)(1-d).
\]
If $d=1$, the constraints
in~\eqref{eq:two-mode-domain} force 
$r_j=r_k=1$, $z=0$, and $s=1$.
The SDP contribution $1-d$ then vanishes, while the
rounded contribution is $\frac{1-\lambda}{2}\ge0$, so last inequality in~\eqref{eq:PSD-termwise-comparison} holds for any $\alpha$. However, having
$(R_{jj},R_{kk},L_\star(n_jn_k),S_{jk},R_{jk})
\in\mathcal D_{jk}$ requires $\alpha\le\beta(\lambda)$
for~\eqref{eq:PSD-termwise-comparison} to hold.

Therefore, for a fixed $\lambda$, all the three inequalities in~\eqref{eq:PSD-termwise-comparison} hold with
the common factor $\alpha=\min\left\{\frac{1+\lambda}{2},\beta(\lambda)\right\}$.
Choosing the best weight is then equivalent to solving
\begin{equation}
\label{eq:PSD-local-optimization}
    \alpha_{\mathrm{PSD}}
    :=\max_{\lambda\in[0,1]}
      \min\left\{\frac{1+\lambda}{2},\,\beta(\lambda)\right\}.
\end{equation}
\begin{lemma}[Optimal PSD weight]
\label{lem:PSD-optimal-comparison} Let $\kappa$ be as in \eqref{eq:PSD-kappa}. Then, 
\begin{equation}
\label{eq:PSD-comparison-weight}
\alpha_{\mathrm{PSD}}=\frac{1+\lambda_{\rm PSD}}{2}
       =\frac{1+4\kappa}{2+4\kappa}\approx0.82699, \quad\text{with}\quad\lambda_{\rm PSD}=\frac{2\kappa}{1+2\kappa}\approx0.65399.
\end{equation}
\end{lemma}
\begin{proof}
At $\lambda=\lambda_{\rm PSD}$, applying Lemma \ref{lem:PSD-occupation-moments} with
$(r_j,r_k,d,s,z)
=(R_{jj},R_{kk},L_\star(n_jn_k),S_{jk},R_{jk})$
and using Lemma \ref{lem:PSD-scalar-comparison} gives
\[
\begin{aligned}
    &\lambda_{\rm PSD}\bigl(1-R_{jj}R_{kk}+|R_{jk}|^2\bigr)
     +\frac{1-\lambda_{\rm PSD}}{2}
        \left(1+\frac{\arccos(S_{jk})}{\pi}\right)\\
    &\quad\ge
    \lambda_{\rm PSD}\left(
    1-L_\star(n_jn_k)-\frac{(1-S_{jk})^2}{16}\right)
    +\frac{1-\lambda_{\rm PSD}}{2}
    \left(1+\frac{\kappa}{4}(1-S_{jk})^2\right)\\
    &\quad=\frac{1+\lambda_{\rm PSD}}{2}
       \bigl(1-L_\star(n_jn_k)\bigr)
       +\frac{1-\lambda_{\rm PSD}}{2}L_\star(n_jn_k)\\
    &\quad\ge\frac{1+\lambda_{\rm PSD}}{2}
       \bigl(1-L_\star(n_jn_k)\bigr).
\end{aligned}
\]
The terms involving $(1-S_{jk})^2$ cancel because
$2\kappa(1-\lambda_{\rm PSD})=\lambda_{\rm PSD}$. Thus $\beta(\lambda_{\rm PSD})\ge\frac{1+\lambda_{\rm PSD}}{2}$, which proves
$\alpha_{\mathrm{PSD}}\ge\frac{1+\lambda_{\rm PSD}}{2}$.

For $\lambda\le\lambda_{\rm PSD}$, we can write 
$$\min\left\{\frac{1+\lambda}{2},\beta(\lambda)\right\}\leq \frac{1+\lambda}{2}\le\frac{1+\lambda_{\rm PSD}}{2}.$$

For $\lambda>\lambda_{\rm PSD}$, let $t_\star\approx0.96881\in]0,1]$ be the minimizer
in~\eqref{eq:PSD-kappa} and consider the following
feasible point in~\eqref{eq:PSD-interaction-optimization}:
\[  
(r_j,r_k,d,s,z)=\left(\frac{t_\star}{2}, \frac{t_\star}{2},0,1-2t_\star,0\right)\in \mathcal{D}_{jk}.
\]
Evaluating the interaction ratio from \eqref{eq:PSD-interaction-optimization} at this point gives
\begin{align*}
    \beta(\lambda)
    \le \lambda\left(1-\frac{t_\star^2}{4}\right)
      +\frac{1-\lambda}{2}
        \left(1+\frac{\arccos(1-2t_\star)}{\pi}\right)&= \lambda\left(1-\frac{t_\star^2}{4}\right)
      +\frac{1-\lambda}{2}\left(1+\kappa t_\star^2\right)\\
    &= \frac{1+\lambda}{2}
      -\frac{(1+2\kappa)t_\star^2}{4}(\lambda-\lambda_{\rm PSD})\\
    &= \frac{1+\lambda_{\rm PSD}}{2}
      +(\lambda-\lambda_{\rm PSD})
        \left(\frac12-\frac{(1+2\kappa)t_\star^2}{4}\right).
\end{align*}
The coefficient of $\lambda-\lambda_{\rm PSD}$ is $\frac12-\frac{(1+2\kappa)t_\star^2}{4}
\approx-0.178132<0$, 
so $\beta(\lambda)\le\frac{1+\lambda_{\rm PSD}}{2}$
for every $\lambda>\lambda_{\rm PSD}$.
We can now state the resulting approximation guarantee.
$\lambda\le\lambda_{\rm PSD}$ yields
$\alpha_{\rm PSD}\le\frac{1+\lambda_{\rm PSD}}{2}$.
\end{proof}
\noindent
Combining these results, we obtain the following approximation guarantee.
\begin{theorem}
\label{thm:PSD-guarantee}
For the Hamiltonian~\eqref{eq:H-PSD}, the state selected
by~\eqref{eq:state-selection} satisfies
\begin{equation}\nonumber
    \mathbb E\!\left[
       \langle H_{\mathrm{PSD}}\rangle_{\psi_{\mathrm{alg}}}\right]
    \ge\frac{1+4\kappa}{2+4\kappa}
       \operatorname{SDP}_2(H_{\mathrm{PSD}})
    \ge\alpha_{\rm PSD}
 \operatorname{OPT}(H_{\mathrm{PSD}}).
\end{equation}
\end{theorem}
\begin{proof}
Take $\lambda=\lambda_{\rm PSD}$ from
Lemma~\ref{lem:PSD-optimal-comparison}.
Multiplying the inequalities
in~\eqref{eq:PSD-termwise-comparison} by the corresponding
nonnegative coefficients of $H_{\mathrm{PSD}}$ and summing,
then using~\eqref{eq:best-candidate-bound}, gives
\[
\mathbb E\!\left[
    \langle H_{\mathrm{PSD}}\rangle_{\psi_{\mathrm{alg}}}
\right]
\ge \lambda_{\rm PSD} E_{\mathrm Q}+(1-\lambda_{\rm PSD})E_{\mathrm{cut}}
\ge \alpha_{\mathrm{PSD}}L_\star(H_{\mathrm{PSD}}).
\]
The result follows from the fact that, by construction, 
$L_\star(H_{\mathrm{PSD}})
=\operatorname{SDP}_2(H_{\mathrm{PSD}})
\ge\OPT(H_{\mathrm{PSD}})$.
\end{proof}
%%%%%%%%%%%%%%%%%%%%%%%%%%%%%%%%%%%%%%%%%%%%%%%%%%%%%%%%%%%%%%%%
\begin{remark}[Signed hopping terms]
\label{rem:PSD-signed-hopping}
The same approximation guarantee holds when the hopping
operators have arbitrary signs. Writing
$q_{jk}:=a_j^\dagger a_k+a_k^\dagger a_j$, consider
\[
\begin{aligned}
    H={}&\sum_{\{j,k\}\in E'}w'_{jk}(\mathbb{I}-s_{jk}q_{jk})+\sum_{\{j,k\}\in E}w_{jk}(\mathbb{I}-n_jn_k)
       +\sum_{j\in V}\mu_jn_j,
\end{aligned}
\]
where $w_{jk}',w_{jk},\mu_j\ge0$ and
$s_{jk}\in\{\pm1\}$. This normalization corresponds to \eqref{eq:H-PSD} when $s_{jk}=1$ for $\{j,k\}\in E'$. Every summand remains positive semidefinite,
since $\mathbb{I}\pm q_{jk}\succeq0$
by~\eqref{eq:PSD-hopping-squares}. The proof of Theorem~\ref{thm:PSD-guarantee} still applies for this Hamiltonian. Indeed,
the hopping SDP value becomes
$1-2s_{jk}\operatorname{Re}R_{jk}\in[0,2]$, while its expected
value in the occupation-basis state is still one.
Thus~\eqref{eq:PSD-hopping-ratio} holds with this substitution as well. Moreover, the Gaussian blending argument of~\cite[Section~5.2]{STH} applies when
$q_{jk}$ is replaced by $s_{jk}q_{jk}$, since
$-\mathbb{I}\preceq s_{jk}q_{jk}\preceq \mathbb{I}$ for any choice of $s_{jk}$.
Thus, for this normalization with arbitrary hopping
    signs, the argument from \cite{STH} still gives the factor $0.637$, while
our rounding schemes ensure $\alpha_{\mathrm{PSD}}\approx0.82699$.
\end{remark}
\subsection{Fermionic Max Cut}
\label{sec:fmc}

Fermionic Max Cut (FMC), introduced in~\cite[Section~5.3]{STH}, is
another model with positive semidefinite edge terms. For a graph
$G=(V,E)$ with $w_{jk}\ge0$ for all $\{j,k\}\in E$, its Hamiltonian is
\begin{equation}
\label{eq:Hfmc}
    \Hf=\sum_{\{j,k\}\in E}w_{jk}h_{jk},
    \quad
    h_{jk}:=\frac12\left(
       n_j+n_k-2n_jn_k-a_j^\dagger a_k-a_k^\dagger a_j
    \right).
\end{equation}
The Hamiltonian $H_{\mathrm{FMC}}$ has the form \eqref{eq: H-generic}, with
\[
    \gamma=0,\quad E'=E,\quad
    \mu_j=\frac12\sum_{k:\{j,k\}\in E}w_{jk},\quad
    w'_{jk}=\frac12w_{jk}.
\]
In terms of the
Majorana operators~\eqref{eq:majorana-definition}, each edge term
can be factored as 
in~\cite[Eq.~(38)]{STH}:
\[
    h_{jk}=
    \left(\frac{\mathbb{I}-i\chi_{2j-1}\chi_{2k}}{2}\right)
    \left(\frac{\mathbb{I}+i\chi_{2j}\chi_{2k-1}}{2}\right)
    \succeq0.
\]
The Majorana CAR \eqref{eq:majorana-CAR}  imply that the two factors are
commuting orthogonal projectors. Their product is therefore
also an orthogonal projector, and hence positive semidefinite, i.e., $h_{jk}\succeq0$ for all $\{j,k\}\in E$. Since $w_{jk}\ge0$,
we conclude that $\Hf\succeq0$.

Notice that the approximation ratio established in Theorem \ref{thm:PSD-guarantee} does not apply for Hamiltonians in \eqref{eq:Hfmc} because of the constant shift term, i.e., 
\[
    H_{\rm PSD} =\Hf+\frac32\left(\sum_{\{j,k\}\in E}w_{jk}\right)\mathbb{I}.
\]
We therefore analyze the approximation ratio for
$H_{\mathrm{FMC}}$ separately, by applying the two rounding schemes
from Section \ref{sec:rounding} to an optimal solution $L_\star\in\mathfrak{L}_2$ of its
level-two fermionic moment relaxation. 

For an edge $\{j,k\}\in E$, by definition, $L_\star(n_j)=R_{jj}$ and
$L_\star(a_j^\dagger a_k)=R_{jk}$, so that evaluating the SDP contribution of the given edge is straightforward:
\begin{equation}
\label{eq:fmc-sdp-RS}
    L_\star(h_{jk})
    =\frac{R_{jj}+R_{kk}}2-L_\star(n_jn_k)
       -\operatorname{Re}R_{jk}.
\end{equation}
Furthermore, substituting the moments from Proposition \ref{prop:SDP-cut-moments} (i)--(ii), we obtain 
\begin{equation}
\label{eq:fmc-cut}
    \mathbb E\!\left[
       \langle h_{jk}\rangle_{\psi_{\mathrm{SDP-cut}}}\right]
    =\frac12\Pr(y_j\ne y_k)
    =\frac{\arccos(S_{jk})}{2\pi}.
\end{equation}
Lastly, Proposition~\ref{prop:SDP-Q-moments} (i)--(ii),
together with~\eqref{eq:fmc-sdp-RS}, gives the expected edge
contribution of the state coming from our first rounding scheme:
\begin{equation}
\label{eq:fmc-Q-exact}
\begin{aligned}
    \mathbb E\!\left[
       \langle h_{jk}\rangle_{\psi_{\mathrm{SDP-Q}}}\right]
    &=L_\star(h_{jk})+L_\star(n_jn_k)
       -R_{jj}R_{kk}+|R_{jk}|^2\\
    &=\frac{R_{jj}+R_{kk}}2-R_{jj}R_{kk}
       +|R_{jk}|^2-\operatorname{Re}R_{jk}.
\end{aligned}
\end{equation}
As in the previous subsection, we balance the strength of each rounding scheme by choosing a common weight
that gives the best uniform guarantee across edges. In other words, the task is to combine
\eqref{eq:fmc-cut} and~\eqref{eq:fmc-Q-exact} so as to retain
the largest uniform fraction of~\eqref{eq:fmc-sdp-RS} over the set of feasible pseudo-moments $\mathcal D_{jk}$ described in Lemma~\ref{lem:two-mode-domain}. This leads to the following optimization problem:
\begin{equation}
\label{eq:fmc-RS-optimization}
\alpha_{\mathrm{FMC}}:=
\max_{\lambda\in[0,1]}
\inf_{\substack{
    (r_j,r_k,d,s,z)\in\mathcal D_{jk},\\
    \frac{r_j+r_k}{2}-d-\operatorname{Re}z>0
}}
\frac{
    \lambda\left(
        \frac{r_j+r_k}{2}-r_jr_k+|z|^2
        -\operatorname{Re}z
    \right)
    +(1-\lambda)\frac{\arccos(s)}{2\pi}
}{
    \frac{r_j+r_k}{2}-d-\operatorname{Re}z
}.
\end{equation}
If $L_\star(h_{jk})=0$, which corresponds to 
$\frac{r_j+r_k}{2}-d-\operatorname{Re}z=0$,
then $h_{jk}\succeq0$ implies $\lambda\,\mathbb E\!\left[
    \langle h_{jk}\rangle_{\psi_{\mathrm{SDP-Q}}}
\right]
+(1-\lambda)\,\mathbb E\!\left[
    \langle h_{jk}\rangle_{\psi_{\mathrm{SDP-cut}}}
\right]
\ge0=\alpha L_\star(h_{jk})$ 
for every $\lambda,\alpha\in[0,1]$. Thus, this case
places no restriction on the approximation ratio.
\begin{lemma}[Optimal FMC weight]
\label{lem:fmc-optimal-comparison}
Let $\theta_0\in\left]\frac\pi2,\pi\right[$ be the unique solution of
\begin{equation}
\label{eq:fmc-angle}
    \pi\sin\theta
       \bigl(1+\cos\theta+\theta\sin\theta\bigr)=2.
\end{equation}
Then the maximum in~\eqref{eq:fmc-RS-optimization} is attained at
\begin{equation}
\label{eq:fmc-analytic-solution}
    \lambda_{\mathrm{FMC}}
       =\frac{2}{2+\pi\theta_0\sin^2\theta_0}\approx0.52237,
    \quad\text{so that}\quad
    \alpha_{\mathrm{FMC}}
       =\frac{1}{\pi\sin\theta_0}\approx0.67971.
\end{equation}
\end{lemma}

\begin{proof}
First, $\theta_0$ exists and is unique. The left-hand side
of~\eqref{eq:fmc-angle} is continuous and strictly decreasing
on $\left[\frac\pi2,\pi\right]$, from
$\pi\left(1+\frac\pi2\right)>2$ to zero (see Figure \ref{fig:fmc-comparison-proof}, left).
It therefore takes the value $2$ at exactly one point
$\theta_0\in\left]\frac\pi2,\pi\right[$.

We will prove that selecting the weight  $\lambda_{\mathrm{FMC}}$ guarantees the ratio of at least 
$\alpha_0:=\frac{1}{\pi\sin\theta_0}$ on every feasible tuple by first reducing the edge analysis to a univariate problem in
$\theta=\arccos(s)$ (\textit{Step I}), and subsequently considering two different possible cases (\textit{Steps II-III} and \textit{Step IV}). We will then explicitly construct a
feasible tuple on which no weight $\lambda$ can give a ratio larger than $\alpha_0$ (\textit{Step V}).

For brevity, given  $(r_j,r_k,d,s,z)\in\mathcal D_{jk}$, let us denote the following values as
\[
\begin{aligned}
    e:=\frac{r_j+r_k}{2}-d-\operatorname{Re}z,\quad
    q:=\frac{r_j+r_k}{2}-r_jr_k+|z|^2
        -\operatorname{Re}z,\quad c:=\frac{\arccos(s)}{2\pi}.
\end{aligned}
\]

\begin{itemize}
\setlength{\itemsep}{0.5em}

\item \textit{Step I: Reducing the edge comparison to $s$ and $e$.} Fix a tuple in $\mathcal D_{jk}$ such that $e>0$.
The value $c$ already depends only on $s$. Let us therefore establish a lower bound for $q$ in terms of $s$ and $e$,
so that the ratio in \eqref{eq:fmc-RS-optimization} depends only on these two quantities.

First, rewrite the constraint $s=1-2r_j-2r_k+4d$ as
$\frac{1-s}{4}=\frac{r_j+r_k}{2}-d$. Combining this identity with the definition of $e$ gives
\begin{equation}
\label{eq: S-L-ReR}
\operatorname{Re}z=\frac{1-s}{4}-e.
\end{equation}
By Lemma~\ref{lem:two-mode-domain}, every tuple in
$\mathcal D_{jk}$ is realized by a two-mode density matrix.
Applying \eqref{eq:PSD-joint-occupation-loss} to its associated pseudo-moment functional gives $r_jr_k-d\le\frac{(1-s)^2}{16}$. 
Using this bound, together with $|z|^2\ge(\operatorname{Re}z)^2$ and \eqref{eq: S-L-ReR}, we obtain
\begin{equation}
\label{eq:fmc-quadratic-comparison}
\begin{aligned}
    q=e-(r_jr_k-d)+|z|^2
    \ge e-\frac{(1-s)^2}{16}
       +(\operatorname{Re}z)^2&=e-\frac{(1-s)^2}{16}
       +\left(\frac{1-s}{4}-e\right)^2\\
    &=e-\frac{1-s}{2}e+e^2=e\left(\frac{1+s}{2}+e\right).
\end{aligned}
\end{equation}
\item \textit{Step II: The case $s\le\frac 12$.} In this case,  $\theta=\arccos(s)\in\left[\frac{\pi}{3},\pi\right]$.
Using the definition of $c$ together with
\eqref{eq:fmc-quadratic-comparison}, we can derive the following lower bound on the approximation ratio:
\[
\begin{aligned}
    \frac{\lambda_{\mathrm{FMC}}q
          +(1-\lambda_{\mathrm{FMC}})c}{e}
    &\ge
       \frac{\lambda_{\mathrm{FMC}}}{2}(1+\cos\theta)
       +\lambda_{\mathrm{FMC}}e
       +\frac{(1-\lambda_{\mathrm{FMC}})\theta}{2\pi e}\\
    &\ge
       \frac{\lambda_{\mathrm{FMC}}}{2}(1+\cos\theta)
       +\sqrt{\frac{
           2\lambda_{\mathrm{FMC}}
           (1-\lambda_{\mathrm{FMC}})\theta}{\pi}}:=F(\theta).
\end{aligned}
\]
The last step follows from the AM-GM inequality. It therefore suffices to show that
$F(\theta)\ge\alpha_0$ for every
$\theta\in[\frac{\pi}{3},\pi]$ (see Figure \ref{fig:fmc-comparison-proof}, right). 
\item \textit{Step III: Minimizing $F$.} The specific weight choice
in equation \eqref{eq:fmc-analytic-solution} gives $\sqrt{\frac{
        2\lambda_{\mathrm{FMC}}
        (1-\lambda_{\mathrm{FMC}})}{\pi}}
    =\lambda_{\mathrm{FMC}}\sqrt{\theta_0}\sin\theta_0$. Thus $F$ and its derivative can be written in a simpler form:
\[
\begin{aligned}
    F(\theta)
    =\frac{\lambda_{\mathrm{FMC}}}{2}
      \left(
          1+\cos\theta
          +2\sin\theta_0\sqrt{\theta_0\theta}
      \right),\quad
    F'(\theta)=\frac{\lambda_{\mathrm{FMC}}}{2\sqrt\theta}
      \left(
          \sqrt{\theta_0}\sin\theta_0
          -\sqrt\theta\sin\theta
      \right).
\end{aligned}
\]
Let us show that $F$ decreases up to $\theta_0$ and
increases afterwards. 
Rearranging~\eqref{eq:fmc-angle}, we can write $ \theta_0\sin^2\theta_0
    =\frac2\pi-\sin\theta_0(1+\cos\theta_0)
    <\frac2\pi$. 
Hence,
\[
\begin{aligned}
    F'\left(\frac \pi3\right)
    =\frac{\lambda_{\mathrm{FMC}}}{2\sqrt{\frac \pi3}}
       \left(
          \sqrt{\theta_0\sin^2\theta_0}
          -\frac{\sqrt\pi}{2}
       \right)<
       \frac{\lambda_{\mathrm{FMC}}}{2\sqrt{\frac \pi3}}
       \left(
          \sqrt{\frac2\pi}-\frac{\sqrt\pi}{2}
       \right)
       <0.
\end{aligned}
\]
Moreover, on $\left[\frac \pi3,\pi\right]$, $F'$ is a sum of the convex
functions $\frac{1}{\sqrt{\theta}}$ and $-\sin\theta$ with positive
coefficients
$\frac12\lambda_{\mathrm{FMC}}\sqrt{\theta_0}\sin\theta_0$
and $\frac12\lambda_{\mathrm{FMC}}$, respectively (recall that $\sin\theta_0>0$ since $\theta_0\in\left]\frac \pi2,\pi\right[$). Thus, $F'$ is convex on $\left[\frac \pi3,\pi\right]$. With that in mind, consider the chord of $F'$ passing through
$\left(\frac \pi3,F'(\frac \pi3)\right)$ and $(\theta_0,0)$.
For $\theta\in\left[\frac \pi3, \theta_0\right[$, $F'$ lies below this chord
and therefore $F'(\theta)<0$. For $\theta=\theta_0$, $F'(\theta_0)=0$. Finally, for $\theta\in]\theta_0,\pi]$,
convexity forces $F'$ to be above the extension of that chord, implying $F'(\theta)>0$. 
Thus, $F$ decreases up to $\theta_0$ and increases afterwards,
so $\displaystyle \min_{\frac \pi3\leq\theta\le\pi}F(\theta)=F(\theta_0)$.

Dividing \eqref{eq:fmc-angle} by
$2\pi\sin\theta_0$ and using $\alpha_0=\frac{1}{\pi\sin\theta_0}$ yields $\frac{1+\cos\theta_0}{2}+\frac{\theta_0\sin\theta_0}{2}=\alpha_0$. 
Therefore,
\[
\begin{aligned}
    \frac{F(\theta_0)}{\lambda_{\mathrm{FMC}}}
    =\frac{1+\cos\theta_0}{2}
       +\theta_0\sin\theta_0=\alpha_0+\frac{\theta_0\sin\theta_0}{2}=\alpha_0
       \left(1+\frac{\pi\theta_0\sin^2\theta_0}{2}\right)
      =\frac{\alpha_0}{\lambda_{\mathrm{FMC}}}.
\end{aligned}
\]
Thus, for all feasible tuples with $s\le\frac12$ and $e>0$,
the approximation ratio at $\lambda_{\rm FMC}$ is at least
$F(\theta_0)=\alpha_0$.

\item \textit{Step IV: The case $s\ge\frac12$.} By~\eqref{eq:fmc-quadratic-comparison}, since $e>0$ and
$s\ge\frac12$,
\[
    \frac{q}{e}
    \ge\frac{1+s}{2}+e
    \ge\frac34.
\]
To bound $\frac{c}{e}$, the $2\times2$ positive semidefinite
constraint defining $\mathcal D_{jk}$ implies
$\det\begin{psmallmatrix}
r_j-d & \overline z\\
z & r_k-d
\end{psmallmatrix}
=(r_j-d)(r_k-d)-|z|^2\ge0$.
Hence, by the AM--GM inequality,
\[
|\operatorname{Re}z|
\le |z|
\le \sqrt{(r_j-d)(r_k-d)}
\le \frac{r_j+r_k-2d}{2}
=\frac{1-s}{4}.
\]
Combining this with~\eqref{eq: S-L-ReR}, we obtain
$0<e=\frac{1-s}{4}-\operatorname{Re}z
    \le\frac{1-s}{2}$. In particular, $s<1$, and therefore
\[
    \frac{c}{e}
    \ge\frac{\arccos(s)}{\pi(1-s)}
    \ge\frac23.
\]
The last inequality follows because $s\mapsto\arccos(s)$ is concave
on $\left[\frac12,1\right]$ and hence lies above the chord
$\frac{2\pi}{3}(1-s)$ joining its endpoint values.

Since $\theta_0\sin^2\theta_0<\frac{2}{\pi}$,
\eqref{eq:fmc-analytic-solution} gives
$\lambda_{\mathrm{FMC}}>\frac12$.
Consequently,
\[
    \frac{\lambda_{\mathrm{FMC}}q
          +(1-\lambda_{\mathrm{FMC}})c}{e}
    \ge\frac34\lambda_{\mathrm{FMC}}
       +\frac23\left(1-\lambda_{\mathrm{FMC}}\right)
    =\frac23+\frac{\lambda_{\mathrm{FMC}}}{12}
    >\frac{17}{24}>\alpha_0.
\]
\noindent
Thus, we deduce that the approximation ratio at $\lambda_{\rm FMC}$ is at least $\alpha_0$ for all feasible tuples with $s\ge\frac12$ and $e>0$. 
\begin{figure}[H]
    \centering \begin{minipage}{\linewidth}
\centering
\definecolor{fmcblue}{HTML}{2563A6}
\definecolor{fmcgreen}{HTML}{43866A}
\definecolor{fmcred}{HTML}{B54D54}
\def\fmctheta{2.654222328018}
\def\fmcalpha{0.679707787414}
\def\fmclambda{0.522370247775}
\pgfplotsset{fmcaxis/.style={
    width=0.475\linewidth,height=6.2cm,
    scale only axis=false,
    axis lines=left,axis line style={black!65},
    tick align=outside,tick style={black!65},
    label style={font=\small},tick label style={font=\small},
    title style={font=\small,yshift=2pt},
    ymajorgrids=true,grid style={black!8},
    clip=false,
}}
\begin{tikzpicture}
\begin{groupplot}[fmcaxis,group style={group size=2 by 1,horizontal sep=1.8cm}]
\nextgroupplot[
    xmin=pi/2,xmax=pi,ymin=0,ymax=9.4,
    xlabel={$\theta$},
    title={(a) The defining equation for $\theta_0$},
    xtick={1.570796327,2.654222328018,3.141592654},
    xticklabels={$\frac{\pi}{2}$,$\theta_0$,$\pi$},
    ytick={0,2,4,6,8},yticklabels={$0$,$2$,$4$,$6$,$8$}]
    \addplot[fmcred,dashed,thick,domain=pi/2:pi] {2};
    \addplot[fmcblue,very thick,domain=pi/2:pi,samples=220]
      {pi*sin(deg(x))*(1+cos(deg(x))+x*sin(deg(x)))};
    \addplot[fmcblue,densely dotted] coordinates
      {(2.654222328018,0)(2.654222328018,2)};
    \addplot[only marks,mark=*,mark size=2.3pt,fmcblue]
      coordinates {(2.654222328018,2)};
    \node[text=fmcblue,font=\footnotesize] at (rel axis cs:0.54,0.93)
      {$\pi\sin\theta\left(1+\cos\theta+\theta\sin\theta\right)$};
    \node[text=fmcred,font=\small,anchor=south east]
      at (axis cs:3.1,2.1) {$y=2$};
\nextgroupplot[
    title={(b) Lower bounds at $\lambda_{\mathrm{FMC}}$},
    xmin=0,xmax=pi,ymin=0.65,ymax=0.85,
    xlabel={$\theta$},
    ylabel={Lower bound on the ratio},
    xtick={0,1.047197551,2.654222328018,3.141592654},
    xticklabels={$0$,$\frac{\pi}{3}$,$\theta_0$,$\pi$},
    ytick={0.679707787414,0.75,0.8,0.85},
    yticklabels={$\alpha_0$,$0.75$,$0.80$,$0.85$}]
    \path[fill=fmcgreen,fill opacity=0.09]
        (axis cs:0,0.708333333) rectangle (axis cs:1.047197551,0.85);
    \addplot[fmcred,dashed,thick,domain=0:pi] {\fmcalpha};
    \addplot[fmcgreen,very thick] coordinates {(0,0.708333333)(1.047197551,0.708333333)};
    \addplot[black!35,densely dotted] coordinates {(1.047197551,0.65)(1.047197551,0.85)};
    \addplot[fmcblue,very thick,domain=pi/3:pi,samples=220]
      {\fmclambda/2*(1+cos(deg(x)))+sqrt(2*\fmclambda*(1-\fmclambda)*x/pi)};
    \addplot[fmcblue,densely dotted] coordinates {(2.654222328018,0.65)(2.654222328018,0.679707787414)};
    \addplot[only marks,mark=*,mark size=2.3pt,fmcblue] coordinates {(2.654222328018,0.679707787414)};
    \node[align=center,text=fmcgreen,font=\small] at (axis cs:0.52,0.804)
      {$s\ge\frac{1}{2}$\\[3pt]\textit{Step IV}};
    \node[text=fmcgreen,font=\small,anchor=south] at (axis cs:0.52,0.713) {$\frac{17}{24}$};
    \node[align=center,text=fmcblue,font=\small] at (axis cs:2.1,0.823)
      {$s\le\frac{1}{2}$\\[3pt]\textit{Steps II--III}};
    \node[text=fmcblue,font=\small] at (axis cs:1.75,0.765) {$F$};
\end{groupplot}
\end{tikzpicture}
\par\vspace{5pt}\captionsetup{font=small}
\captionof{figure}{%
\textbf{Left}: the unique solution $\theta_0\in\left]\frac \pi2,\pi\right[$
of $\pi\sin\theta(1+\cos\theta+\theta\sin\theta)=2$.
\textbf{Right}: the lower bounds established in \textit{Steps II--IV}
at $\lambda=\lambda_{\mathrm{FMC}}$.
For $\cos{\theta=}s\le\frac12$, the bound $F(\theta)$ attains
its minimum $\alpha_0$ at $\theta_0$, whereas
for $s\ge\frac12$, the ratio is greater than
$\frac{17}{24}$.
}
\label{fig:fmc-comparison-proof}
\end{minipage}
\end{figure}
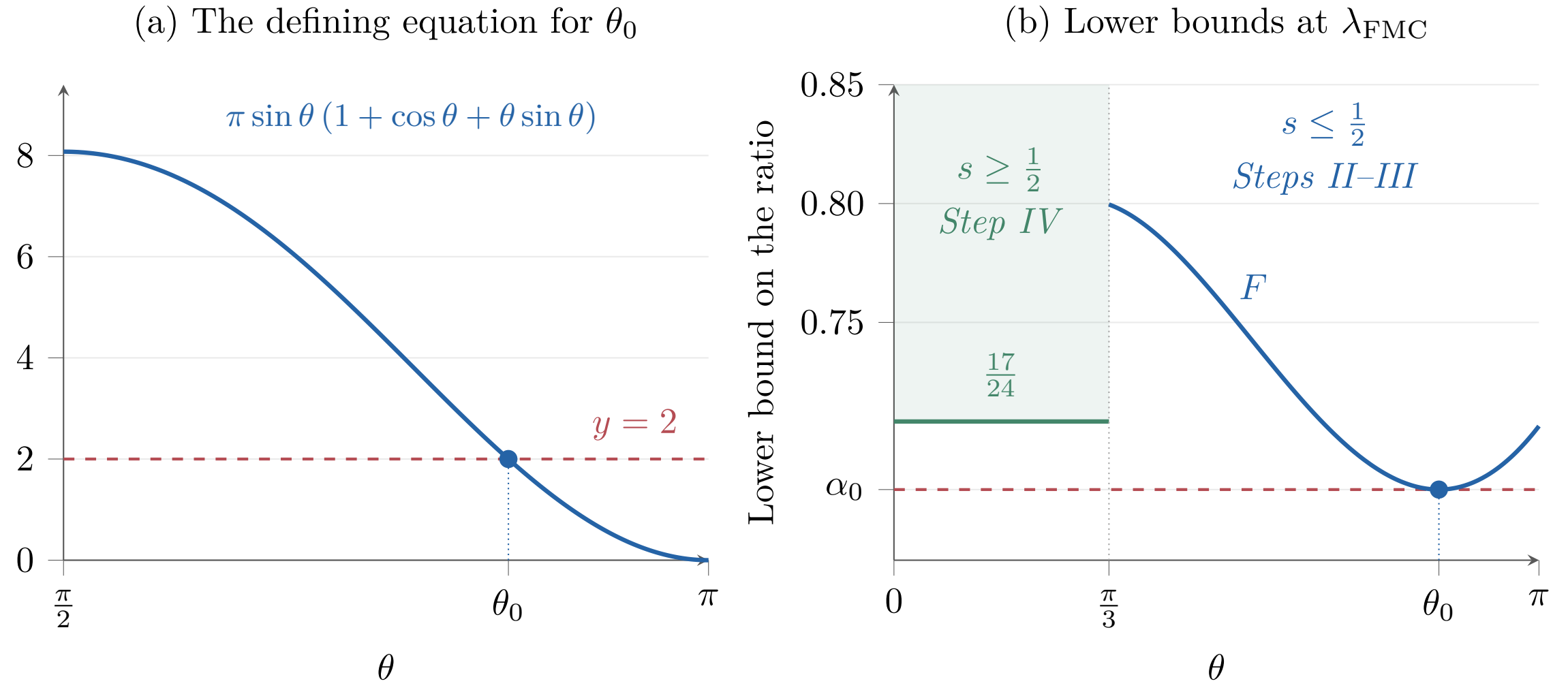
\item \textit{Step V: Bound attainment.} 
To prove optimality of $\lambda_{\rm FMC}$, we construct a feasible tuple in
$\mathcal D_{jk}$ satisfying
$\frac{q}{e}=\frac{c}{e}=\alpha_0$, implying that  
$\frac{\lambda q+(1-\lambda)c}{e}=\alpha_0=F(\theta_0)$ for every $\lambda\in[0,1]$.
Set
\[
\begin{aligned}
 r_j=r_k&=\frac{1-\cos\theta_0}{4},\quad
    z=\frac{1-\cos\theta_0}{4}
       -\frac{\theta_0\sin\theta_0}{2},\quad
    d=0,\quad s=\cos\theta_0.
\end{aligned}
\]
The tuple above is selected to attain equality in the
bounds from \textit{Steps I--II} that hold at $\lambda=\lambda_{\mathrm{FMC}}$.
Indeed, the choices of $r_j,r_k,d$ and $z$ make
\eqref{eq:fmc-quadratic-comparison} an equality.
Moreover, for this tuple, $e=r_j-z=\frac{\theta_0\sin\theta_0}{2}$, 
which also turns the AM--GM inequality of \textit{Step II} into
an equality.
Consequently,
\[
    \frac{q}{e}
    =\frac{1+\cos\theta_0+\theta_0\sin\theta_0}{2},
    \qquad
    \frac{c}{e}=\frac{1}{\pi\sin\theta_0},
\]
and, thanks to \eqref{eq:fmc-angle}, these 
two ratios are equal to $\alpha_0$. It remains to verify that the chosen coordinates are indeed feasible.

Since $d=0$ and $0\le r_j=r_k\le\frac12$,
we have $d\ge0$ and $1-r_j-r_k+d\ge0$.
The identity $s=1-2r_j-2r_k+4d$ also holds
by construction.
It remains to verify the $2\times2$ matrix inequality
in~\eqref{eq:two-mode-domain}.
Since $r_j=r_k$ and $z$ is real, the matrix
has eigenvalues $r_j+z$ and $r_j-z$.
By~\eqref{eq:fmc-angle},
$r_j+z=1-\alpha_0\ge0$, while
$r_j-z=\frac{\theta_0\sin\theta_0}{2}>0$
holds by construction.
Thus, the matrix is positive semidefinite,
and the tuple belongs to $\mathcal D_{jk}$. 
Thus, for every weight $\lambda\in[0,1]$, the infimum
in~\eqref{eq:fmc-RS-optimization} is at most $\alpha_0$.
\end{itemize}
To sum up, \textit{Steps II--IV} show that choosing
$\lambda=\lambda_{\mathrm{FMC}}$ guarantees a ratio of at least
$\alpha_0$ for every feasible tuple with $e>0$, so
$\alpha_{\mathrm{FMC}}\ge\alpha_0$.
\textit{Step V} then provides a feasible tuple whose ratio equals $\alpha_0$
for every $\lambda\in[0,1]$, so
$\alpha_{\mathrm{FMC}}\le\alpha_0$.
Therefore, $\alpha_{\mathrm{FMC}}=\alpha_0$.
\end{proof}
It is now straightforward to prove the following approximation guarantee.
\begin{theorem}
\label{thm:fmc}
For $\Hf$ in~\eqref{eq:Hfmc}, the state $|\psi\rangle_{\rm alg}$ from \eqref{eq:state-selection} satisfies
\begin{equation}\nonumber
    \mathbb E\!\left[
       \langle\Hf\rangle_{\psi_{\mathrm{alg}}}\right]
    \ge\alpha_{\mathrm{FMC}}\operatorname{SDP}_2(\Hf)
    \ge\alpha_{\mathrm{FMC}}\operatorname{OPT}(\Hf).
\end{equation}
\end{theorem}
\begin{proof}
By Lemma~\ref{lem:two-mode-domain}, we have $\in (R_{jj},R_{kk},L_\star(n_jn_k),S_{jk},R_{jk})\in \mathcal D_{jk}$ for every edge $\{j,k\}\in E$.
Lemma~\ref{lem:fmc-optimal-comparison}, together with the
zero-edge-value case discussed above, therefore gives
\[
    \lambda_{\mathrm{FMC}}\mathbb E\!\left[
       \langle h_{jk}\rangle_{\psi_{\mathrm{SDP-Q}}}\right]
    +(1-\lambda_{\mathrm{FMC}})\mathbb E\!\left[
       \langle h_{jk}\rangle_{\psi_{\mathrm{SDP-cut}}}\right]
    \ge\alpha_{\mathrm{FMC}}L_\star(h_{jk}).
\]
Multiplying both sides of this inequality by $w_{jk}\ge0$ and summing over the edges, we obtain
\[
    \lambda_{\mathrm{FMC}}E_{\mathrm Q}
    +(1-\lambda_{\mathrm{FMC}})E_{\mathrm{cut}}
    \ge\alpha_{\mathrm{FMC}}L_\star(\Hf).
\]
The conclusion follows from~\eqref{eq:best-candidate-bound}
and the fact that $L_\star(\Hf)=\operatorname{SDP}_2(\Hf)
    \ge\operatorname{OPT}(\Hf)$. 
\end{proof}
%%%%%%%%%%%%%%%%%%%%%%%%%%%%%%%%%
\section{Approximation guarantees for traceless Hamiltonians}
\label{sec:traceless-guarantees}
We now consider the traceless normalization of classically
interacting fermionic Hamiltonians, following
\cite[Problem~2.1]{STH}:
\begin{equation}
\label{eq:H-traceless}
\begin{aligned}
    H_{\mathrm{tr}}:={}&-\sum_{\{j,k\}\in E'}w'_{jk}
       \bigl(a_j^\dagger a_k+a_k^\dagger a_j\bigr)+
    \sum_{\{j,k\}\in E}w_{jk}\left(\frac14\mathbb{I}-n_jn_k\right)
    +\sum_{j\in V}\mu_j\left(n_j-\frac12\mathbb{I}\right),
\end{aligned}
\end{equation}
where $w_{jk}\ge0$ and $w'_{jk},\mu_j\in\mathbb R$. This is a Hamiltonian of the form~\eqref{eq: H-generic},
    with $\gamma=\frac W4-\frac12\sum_j\mu_j$, where we write $W:=\sum_{\{j,k\}\in E}w_{jk}$ for the total interaction weight.
Each summand has zero trace, so the completely mixed state $\rho_{\rm mix}:=\frac{\mathbb{I}}{2^n}$ satisfies $\langle H_{\rm tr}\rangle_{\rho_{\rm mix}}=0$ and therefore $\operatorname{OPT}(H_{\mathrm{tr}})\ge0$.
Here again, we study the optimization problem from  \eqref{eq: OPTProb}, with
$H=H_{\mathrm{tr}}$.

The centered terms in~\eqref{eq:H-traceless} can have negative
expectations, so the term-wise multiplicative comparisons of
Section~\ref{sec:PSD-guarantees} do not directly apply.
We instead bound the total energy
lost in rounding by an absolute additive amount for Hamiltonians with general interaction graphs in Subsection~\ref{sec: generaltraceless}, and then use these bounds in Subsection \ref{sec:traceless-bipartite}
to derive multiplicative approximation ratios for
Hamiltonians with bipartite interaction graphs.
\subsection{General interaction graphs}\label{sec: generaltraceless}
We start by considering the general case where no restrictions are imposed on the interaction graph $G=(V,E)$.  Throughout this subsection, $L_\star$ is an optimal level-two
functional for $H_{\mathrm{tr}}$, and $R$, $S$, $E_{\mathrm Q}$,
and $E_{\mathrm{cut}}$ are the corresponding pseudo-moments and expected
state energies, as defined in  Section~\ref{sec:rounding}.

The following result shows that the loss incurred by rounding the relaxation solution to the Gaussian state $|\psi_{\rm SDP-Q}\rangle$ is at most $\frac W4$.
\begin{theorem}
\label{thm:traceless-additive}
For $H_{\mathrm{tr}}$ in~\eqref{eq:H-traceless}, construct $|\psi_{\mathrm{SDP-Q}}\rangle$
from $L_\star$ as in~\eqref{eq:SDP-Q-state}. Then,
\begin{equation}
\label{eq:traceless-uniform-bound}
    \mathbb E[\langle H_{\mathrm{tr}}\rangle_{\psi_{\mathrm{SDP-Q}}}]
    \ge\operatorname{OPT}(H_{\mathrm{tr}})-\frac W4.
\end{equation}
\end{theorem}

\begin{proof}
Applying 
Proposition~\ref{prop:SDP-Q-moments} and
equation~\eqref{eq:SDP-Q-energy-difference} to $|\psi_{\mathrm{SDP-Q}}\rangle$ gives
\[
    \mathbb E[\langle H_{\mathrm{tr}}\rangle_{\psi_{\mathrm{SDP-Q}}}]=E_{\mathrm Q}
    =L_\star(H_{\mathrm{tr}})
      -\sum_{\{j,k\}\in E}w_{jk}
       \bigl(R_{jj}R_{kk}-|R_{jk}|^2-L_\star(n_jn_k)\bigr).
\]
By Lemma~\ref{lem:PSD-occupation-moments}, every feasible
level-two functional satisfies $R_{jj}R_{kk}-|R_{jk}|^2-L_\star(n_jn_k)
    \le\frac{(1-S_{jk})^2}{16}\le\frac14$, where the last inequality follows from $S_{jk}\in[-1,1]$.
Therefore, since the weights $w_{jk}$ are nonnegative, applying this
bound to each edge gives
\begin{equation}
\label{eq:traceless-Q-bound}
    \mathbb E[\langle H_{\mathrm{tr}}\rangle_{\psi_{\mathrm{SDP-Q}}}]
    \ge \operatorname{SDP}_2(H_{\mathrm{tr}})-\frac W4
    \ge \operatorname{OPT}(H_{\mathrm{tr}})-\frac W4,
\end{equation}
which is the desired result.
\end{proof}
The bounds above are additive, which naturally raises
the question of whether a constant multiplicative
approximation ratio is possible. In what follows, we argue that the
absence of such a multiplicative guarantee is not merely a limitation
of our rounding schemes, but reflects an inherent
hardness of the general traceless problem under the randomized
Unique Games Conjecture. To see so, let us write 
\begin{equation}
\label{eq:traceless-decomposition}
    H_{\mathrm{tr}}=-\frac14\sum_{\{j,k\}\in E}w_{jk}Z_jZ_k,
\end{equation}
which is a special case of \eqref{eq:H-traceless} with $w'_{jk}=0$ for all $\{j,k\}\in E'$ and
$\mu_j=\frac12\sum_{k:\{j,k\}\in E}w_{jk}$ for all $j\in V$.
We now illustrate that this special case is hard
to approximate within any constant factor. 

For every $x\in\{0,1\}^n$, writing
$\operatorname{Cut}_w(x):=
\sum_{\{j,k\}\in E}w_{jk}\mathbf1_{\{x_j\ne x_k\}}$,
we obtain
\begin{equation}
\label{eq:traceless-maxcut-gain}
    \begin{aligned}
    % H_{\mathrm{tr}}=-\frac14\sum_{\{j,k\}\in E}w_{jk}Z_jZ_k,
    % \quad
    \langle x| H_{\mathrm{tr}}|x\rangle=\frac12\left(\operatorname{Cut}_w(x)-\frac W2\right).
    \end{aligned}
\end{equation}
Recall that, for a weighted graph, {Max-Cut-Gain} asks for a cut that
approximates the maximum gain over the random-cut baseline $\frac W2$, i.e., the gain of a cut $x$ is
$G_w(x):=\operatorname{Cut}_w(x)-\frac W2$.
By~\eqref{eq:traceless-maxcut-gain}, $\langle x| H_{\mathrm{tr}}|x\rangle=\frac{G_w(x)}{2}$, meaning that every state’s $|x\rangle$ energy is exactly half the gain of the
corresponding cut.

Since this specific Hamiltonian is diagonal, there exists an occupation-basis state achieving its largest eigenvalue,  implying that $\OPT(H_{\rm tr})=\frac12\max_{x\in\{0,1\}^n}G_w(x)$. Let $x\in\{0,1\}^n$ denote the random occupation string
obtained by measuring the output state $|\psi_{\rm alg}\rangle$, viewed as a cut. Thus, a constant expected $\alpha$-approximation for the energy would imply that
\[
\begin{aligned}
    \mathbb E[G_w(x)]
    =2\,\mathbb E[\langle H_{\mathrm{tr}}\rangle_{\psi_{\mathrm{alg}}}]\ge2\alpha\,\operatorname{OPT}(H_{\mathrm{tr}})
     =\alpha\max_{x\in\{0,1\}^n}G_w(x).
\end{aligned}
\]
In other words, the
same constant approximation factor $\alpha$ would also hold for Max-Cut-Gain. Under the randomized version of the Unique Games Conjecture  \cite[Theorem~4.3]{KhotODonnell}, this is ruled out for every fixed $\alpha>0$. This obstruction
applies regardless of how the two rounded states are mixed or
selected.

Stronger guarantees become possible with additional
structure. For example, for bipartite graphs, \cite[Theorem~2.3]{STH} provides a polynomial-time constructable Gaussian states with $\frac13$-approximation ratio.
Therefore, in the next subsection, we extend our analysis to bipartite graphs and show that it yields an improved ratio.

\subsection{Bipartite interaction graphs}
\label{sec:traceless-bipartite}
Let $V=A\sqcup B$ be a bipartition of the graph $G=(V,E)$ and
consider the two occupation-basis states
\begin{equation}
\label{eq: bipartitionstates}     |\psi_A\rangle:=
    \prod_{j\in A}a_j^\dagger|\mathrm{vac}\rangle,
    \quad
    |\psi_B\rangle:=
    \prod_{j\in B}a_j^\dagger|\mathrm{vac}\rangle.
\end{equation}
The first state occupies all modes in $A$ and leaves
those in $B$ empty, while the second reverses
these occupations. Both are Slater determinants of
the form~\eqref{eq:mode-occupation-state} with $U=\mathbb{I}$,
and hence pure Gaussian states by
Remark~\ref{rem:Slater}.

Such bipartition states were also considered
in~\cite{STH} in the context of particle-number
constraints. In what follows, we consider a modified algorithm that augments our two
SDP-based rounding schemes with $|\psi_A\rangle$ and
$|\psi_B\rangle$, and returns the state achieving the
largest energy among the four candidates.
We denote the resulting state by
$|\psi_{\mathrm{alg,bip}}\rangle$.
More formally, the modified algorithm outputs
\begin{equation}
\label{eq:state-selection-bipartite}
|\psi_{\mathrm{alg,bip}}\rangle:=
\begin{cases}
|\psi_{\mathrm{alg}}\rangle,
& \text{if }\langle H_{\mathrm{tr}}\rangle_{\psi_{\mathrm{alg}}}
  \ge \max\left\{
  \langle H_{\mathrm{tr}}\rangle_{\psi_A},
  \langle H_{\mathrm{tr}}\rangle_{\psi_B}\right\},\\
|\psi_A\rangle,
& \text{if }\langle H_{\mathrm{tr}}\rangle_{\psi_A}
  \ge \max\left\{
  \langle H_{\mathrm{tr}}\rangle_{\psi_{\mathrm{alg}}},
  \langle H_{\mathrm{tr}}\rangle_{\psi_B}\right\},\\
|\psi_B\rangle,
& \text{otherwise}.
\end{cases}
\end{equation}
\begin{comment}
    Then the modified
randomized polynomial-time algorithm described above produces a Slater determinant 
\end{comment}
\begin{theorem}
\label{thm:traceless-bipartite}
For $H_{\mathrm{tr}}$ in~\eqref{eq:H-traceless},
suppose that the weighted interaction graph
$G=(V,E)$ is bipartite. Then, one can construct a Slater determinant 
$|\psi_{\mathrm{alg,bip}}\rangle$
satisfying
\begin{equation}\nonumber
    \mathbb E[\langle H_{\mathrm{tr}}\rangle_{\psi_{\mathrm{alg,bip}}}]
    \ge\frac12\operatorname{SDP}_2(H_{\mathrm{tr}})
    \ge\frac12\operatorname{OPT}(H_{\mathrm{tr}}).
\end{equation}
\end{theorem}
\begin{proof} Using the properties of Hamiltonian edge terms, it is easy to show that 
\[
\begin{aligned}
    \langle H_{\mathrm{tr}}\rangle_{\psi_A}
    =\frac W4+\frac12
       \left(\sum_{j\in A}\mu_j-\sum_{j\in B}\mu_j\right),\quad
    \langle H_{\mathrm{tr}}\rangle_{\psi_B}
    =\frac W4-\frac12
       \left(\sum_{j\in A}\mu_j-\sum_{j\in B}\mu_j\right).
\end{aligned}
\]
The larger energy being at least as large as the average, we get  $ E_{\mathrm{bip}}:=
    \max\left\{\langle H_{\mathrm{tr}}\rangle_{\psi_A},
                \langle H_{\mathrm{tr}}\rangle_{\psi_B}\right\}
    \ge\frac W4$. 
Combining this with the bound 
$E_{\mathrm Q}\ge\operatorname{SDP}_2(H_{\mathrm{tr}})-\frac W4$ established in Theorem \ref{thm:traceless-additive}, 
and recalling that the algorithm selects the best of
the four candidates, we obtain
\[
\begin{aligned}
    \mathbb E[\langle H_{\mathrm{tr}}\rangle_{\psi_{\mathrm{alg,bip}}}]
    \ge\max\left\{
       \operatorname{SDP}_2(H_{\mathrm{tr}})-\frac W4,\,
       \frac W4\right\}\ge\frac12\operatorname{SDP}_2(H_{\mathrm{tr}})\geq\frac12\OPT(H_{\rm tr}),
\end{aligned}
\]
where the penultimate inequality uses that the maximum of two numbers is
at least their average. \\
Finally, notice that, by construction and Section~\ref{sec:rounding},
all four candidates are Slater determinants.
\begin{comment}
    A bipartition can be found in polynomial time,
and the energies of the two additional occupation-basis
states can be evaluated using the explicit expressions
above. Together with the randomized polynomial-time
SDP-based constructions of Section~\ref{sec:rounding},
this establishes the claimed complexity.
\end{comment}
\end{proof}
\begin{remark}
    Although $|\psi_{\rm SDP-cut}\rangle$ is not needed to
establish the approximation guarantee in
Theorem~\ref{thm:traceless-bipartite}, it may still
improve the algorithm's performance on individual
instances. We therefore retain it as an additional
candidate, since its construction requires little
additional computation once the level-two SDP
relaxation has been solved.
\end{remark}

%%%%%%%%%%%%%%%%%%%%%%%%%%%%%%%%%%%%%%%%%%%%%%%%%%%%%%%%%%%%%%%%%%%%%%%%%%%%%%%
\subsection{Particle-number constraints}
\label{sub: particle-number-constraints}
So far, we have imposed no restrictions on the particle
number, allowing the optimal fermionic state to have
an arbitrary number of occupied modes.
We now consider particle-number constraints, starting
with the relaxed setting of~\cite[Problem~2.6]{STH},
where only the expected number of fermions is prescribed. The case of an exact particle-number constraint
will be considered subsequently. 
We consider the Hamiltonian
\begin{equation}
\label{eq:H-particle-number}
    H=
       -\sum_{\{j,k\}\in E'}w'_{jk}
        \bigl(a_j^\dagger a_k+a_k^\dagger a_j\bigr)+\sum_{\{j,k\}\in E}w_{jk}\left(\frac14\mathbb{I}-n_jn_k\right),
\end{equation}
where $w_{jk}\ge0$ and $w'_{jk}\in\mathbb R$.
This is the model~\eqref{eq:H-traceless} with $\mu_j=0$,
or equivalently~\eqref{eq: H-generic} with
$\gamma=\frac W4$ and $\mu_j=0$ for all $j\in V$. 
\paragraph{Average $q$-particle constraint.} Given an integer $q\in\{0,\ldots,\left\lfloor \frac n2\right\rfloor\}$, the goal is to approximate with fermionic Gaussian states the following optimization problem:
\begin{equation}
\label{eq:particle-number-optimum}
\operatorname{OPT}_{\langle q\rangle}(H)
:=\max_{\rho\in\mathbb{C}^{2^n\times 2^n}}\left\{\operatorname{Tr}(\rho H)\;\mid\;
        \rho\succeq0,\;\ \operatorname{Tr}(\rho)=1,\;\langle N\rangle_\rho=q\right\}.
\end{equation}
Here $N=\sum_{j\in V}n_j$ is the particle-number
operator introduced in Subsection \ref{sec:gaussian}. 
As in Subsection~\ref{sec:hierarchy}, we can relax this
problem using fermionic pseudo-moments, with the constraint
$\langle N\rangle_\rho=q$ translated into the
linear constraint $L(N)=q$ on the pseudo-moment functional.
The minimal level-two fermionic moment relaxation
is therefore
\begin{equation}
\label{eq:particle-number-SDP}
    \operatorname{SDP}_{2,\langle q\rangle}(H)    :=\max_{\substack{L\in\mathfrak L_2,\\L(N)=q}}L(H)
    \ge\operatorname{OPT}_{\langle q\rangle}(H).
\end{equation}
\noindent
Let $L_\star$ be an optimal solution of
\eqref{eq:particle-number-SDP} and set
$R_{jk}=L_\star(a_j^\dagger a_k)$ as before.
Although our first rounding scheme preserves
the particle number in expectation, a single
sampled state does not need to have exactly $q$ particles.
We therefore consider the mixed Gaussian state
$\rho_R$ from~\eqref{eq:gaussian-realization},
which, recalling \eqref{eq:SDP-Q-average} and \eqref{eq:SDP-Q-particle-number}, satisfies
\[
\begin{aligned}
\operatorname{Tr}(\rho_R N)=\E[\Tr(|\psi_{\rm SDP-Q}\rangle
  \langle\psi_{\rm SDP-Q}|N)]=\E[\langle N\rangle_{\psi_{\rm SDP-Q}}]
=\operatorname{Tr}(R)
=L_\star(N)=q.
\end{aligned}
\]
\noindent
Thus, \(\rho_R\) is feasible in \eqref{eq:particle-number-optimum}. By~\eqref{eq:SDP-Q-average} and linearity of the trace,
we also have
\[
   \operatorname{Tr}(\rho_R H)
    =\mathbb E\!\left[
        \langle H\rangle_{\psi_{\mathrm{SDP-Q}}}
      \right]
    =E_{\mathrm Q}.
\]
The argument used to prove Theorem~\ref{thm:traceless-additive}
also applies to the particle-number-constrained relaxation,
since the additional constraint $L(N)=q$ does not affect
the bounds of Lemma~\ref{lem:PSD-occupation-moments}.
Consequently,
\begin{equation}
\label{eq:particle-number-additive}
    \operatorname{Tr}(\rho_R H)=E_{\mathrm Q}
    \ge \operatorname{SDP}_{2,\langle q\rangle}(H)-\frac W4.
\end{equation}
Moreover, this bound  requires no assumption on the interaction graph.  However, whenever the interaction graph is bipartite, it is straightforward to construct a complementary feasible Gaussian
state whose energy is $\frac W4$. Indeed, let $V=A\sqcup B$, with $|A|\ge|B|$. Since $q\le\left\lfloor \frac n2\right\rfloor$,
we can choose a subset $C\subseteq A$ with $|C|=q$ and form the Slater determinant
$\displaystyle|\psi_C\rangle:=
    \prod_{j\in C}a_j^\dagger|\mathrm{vac}\rangle$. Set $\rho_C:=|\psi_C\rangle\langle\psi_C|$ and consider the approximation algorithm that outputs \begin{equation}
\label{eq:state-selection-mean-q}
\rho_{\mathrm{alg},\langle q\rangle}:=
\begin{cases}
\rho_R,
& \text{if }\langle H\rangle_{\rho_R}
  \ge\langle H\rangle_{\rho_C},\\
\rho_C,
& \text{otherwise}.
\end{cases}
\end{equation}
\begin{theorem}
\label{thm:particle-number-bipartite}
For $H$ in~\eqref{eq:H-particle-number}, suppose that $(V,E)$
is bipartite and let $q\in \mathbb{N}$ satisfy $0\le q\le\left\lfloor \frac n2\right\rfloor$. 
Then, the algorithm from \eqref{eq:state-selection-mean-q} outputs
a Gaussian state $\rho_{\mathrm{alg},\langle q\rangle}$ satisfying
$\langle N\rangle_{\rho_{\mathrm{alg},\langle q\rangle}}=q$ and
\begin{equation}
\label{eq:particle-number-bipartite-guarantee}
    \langle H\rangle_{\rho_{\mathrm{alg},\langle q\rangle}}
    \ge\frac12\operatorname{SDP}_{2,\langle q\rangle}(H)
    \ge\frac12\operatorname{OPT}_{\langle q\rangle}(H).
\end{equation}
\end{theorem}

\begin{proof}
By substitution, it is straightforward to see that
\[
    \langle H\rangle_{\rho_C}
    =\sum_{\{j,k\}\in E}w_{jk}\left(\frac14-0\right)-0
    =\frac W4.
\]
Combining this value with~\eqref{eq:particle-number-additive}
and selecting the better state gives
\[
    \langle H\rangle_{\rho_{\mathrm{alg},\langle q\rangle}}
    \ge\max\left\{\operatorname{SDP}_{2,\langle q\rangle}(H)
                   -\frac W4,\,\frac W4\right\}
    \ge\frac12\operatorname{SDP}_{2,\langle q\rangle}(H)\geq\frac12\operatorname{OPT}_{\langle q\rangle}(H),
\]
where the second inequality follows because the
maximum of two numbers is at least their average.
Finally, by construction, both candidates are Gaussian states satisfying the
mean particle-number constraint.
\begin{comment}
    Their construction and energy evaluation require only
solving a level-two SDP, matrix diagonalization, and
finding a graph bipartition, all of which can be
performed in polynomial time to a prescribed numerical
accuracy.
\end{comment}
\end{proof}
For the same mean-particle-number problem on bipartite interaction graphs,
\cite[Theorem~2.7]{STH} gives the factor
$\displaystyle \frac{1}{5-4\frac qn}\in\left[\frac 15,\frac 13\right]$. In contrast, our SDP-based analysis shows that this constant approximation factor is at least $\frac 12$.
\paragraph{Exact $q$-particle constraint.} 
While the preceding result fixes the particle number
only in expectation, many physical applications,
particularly in quantum chemistry, require states
with \textit{exactly} $q$ fermions.
In the discussion following Theorem~2.7 of~\cite{STH},
the authors leave open whether a constant-factor
Gaussian approximation can be obtained under this
stronger constraint. More precisely, this amounts
to considering the exact-particle-number
optimization problem
\begin{align}\label{eq:exact-q-opt}
    \operatorname{OPT}_q(H)
    :=\max_{|\psi\rangle\in\mathbb{C}^{2^n}}\left\{\langle H \rangle_{\psi}\;\mid\;
\langle\psi|\psi\rangle=1,\
       N|\psi\rangle=q|\psi\rangle\right\}
    \le\operatorname{OPT}_{\langle q\rangle}(H),
\end{align}
with $H$ defined in \eqref{eq:H-particle-number}. 

In what follows, we show that the same
$\frac12$-approximation guarantee can be obtained
with exactly $q$ particles on bipartite interaction
graphs, for any integer $0\le q\le\left\lfloor \frac n2\right]$,
using a \textit{two-step rounding} procedure.

In the first step, we relax the exact particle-number
constraint to its mean counterpart and apply the
algorithm of Theorem~\ref{thm:particle-number-bipartite}.
This yields a Gaussian state
$\rho_{\mathrm{alg},\langle q\rangle}$
with mean particle number $q$ and the corresponding
$\frac12$-approximation guarantee.

The second step uses a pairwise rounding procedure
for the occupation probabilities, closely related
to deterministic pipage rounding~\cite{AgeevSviridenko2004}
and to the constructive proof of Lieb's variational
principle for Hartree--Fock theory~\cite{Lieb1981,Bach2022}.  Specifically, if the selected state
is $\rho_{\mathrm{alg},\langle q\rangle}=\rho_R$,
we adjust pairs of fractional eigenvalues of its
one-particle correlation matrix $R$ until all eigenvalues
belong to $\{0,1\}$, while preserving their sum $q$
and without decreasing the energy. This produces a
Slater determinant with exactly $q$ particles, while
retaining the same approximation guarantee. The
following lemma makes this second step explicit.
\begin{lemma}
\label{lem:fixed-particle-rounding}
Let $H$ be as in~\eqref{eq:H-particle-number},
and let $R\in\mathbb C^{n\times n}$ be Hermitian, with $0\preceq R\preceq \mathbb{I}$ and 
$\operatorname{Tr}(R)=q$ for some integer $q\in\mathbb{N}$. Let $\rho_R$ be the mixed Gaussian state associated
with $R$, defined in~\eqref{eq:gaussian-realization}.
Then, one can construct
a Slater determinant $|\psi_{{\rm alg},q}\rangle$ such that
\[
    N|\psi_{{\rm alg},q}\rangle=q|\psi_{{\rm alg},q}\rangle,
    \quad\text{and}\quad
    \langle H\rangle_{\psi_{{\rm alg},q}}\ge\langle H\rangle_{\rho_R}.
\]
\end{lemma}
\begin{proof}

We start by diagonalizing
$R=U\operatorname{diag}(\nu_1,\ldots,\nu_n)U^\dagger$
as in~\eqref{eq:R-eigendecomposition}, and construct
the rotated modes $b_\ell$ as in~\eqref{eq:normal-modes}.
Since $0\preceq R\preceq \mathbb{I}$ and $\operatorname{Tr}(R)=q$,
we have $0\le\nu_\ell\le1$ and
$\sum_{\ell=1}^n\nu_\ell=q$.
In the Gaussian state $\rho_R$, each $\nu_\ell$
represents the probability that mode $b_\ell$
is occupied. We now show how to round the fractional eigenvalues
$\nu_\ell$ to zeros and ones, while preserving their
sum and without decreasing the energy. Since their sum is an integer, whenever fractional
eigenvalues remain, there must be at least two of them.
We repeatedly apply the following two steps until
all eigenvalues are either zero or one.
\begin{enumerate}
    \item \textit{Step I.} Choose any two indices $\ell\ne m$ such that
$\nu_\ell,\nu_m\in]0,1[$, and set
$s:=\nu_\ell+\nu_m$. 
Depending on whether $s\le1$ or $s>1$,
we consider the following two possible replacements
$(\widetilde{\nu}_\ell,\widetilde{\nu}_m)$
of the pair $(\nu_\ell,\nu_m)$:

    \[
    \begin{array}{c|cc}
        &\text{first replacement}&\text{second replacement}\\ \hline
        \text{if}\:s\le1 &(s,0)&(0,s)\\
        \text{if}\:s>1 &(1,s-1)&(s-1,1)
    \end{array}
    \]
    Importantly, either replacement preserves the sum $s$, but makes at least one of
    the two entries equal to zero or to one. We emphasize that, at this stage, all other entries are kept fixed.

    \item \textit{Step II.} 
Form the matrix $\widetilde R
    =U\operatorname{diag}
    (\nu_1,\dots,\widetilde{\nu}_\ell,\dots,
    \widetilde{\nu}_m,\dots,\nu_n)U^\dagger$ for each of the two replacements, 
and evaluate the energy of its associated Gaussian
state using~\eqref{eq:gaussian-energy}.
Keep the replacement yielding the larger energy. If fractional eigenvalues remain, repeat \textit{Step I}
using the updated eigenvalues.
\end{enumerate}
Thus, each iteration makes at least one additional
eigenvalue equal to zero or one, which is then kept
fixed. Since the sum of all eigenvalues remains the
integer $q$, the last two fractional eigenvalues
must sum to one and are therefore both rounded
in the final iteration. Consequently, the procedure
terminates after at most $n-1$ iterations. 
The final eigenvalues
$\widetilde{\nu}_1,\ldots,\widetilde{\nu}_n$
belong to $\{0,1\}$ and sum to $q$.
They therefore define the occupation vector
$\widetilde{x}:=(\widetilde{\nu}_1,\ldots,
\widetilde{\nu}_n)\in\{0,1\}^n$
and the corresponding Slater determinant
\[
    |\psi_{{\rm alg},q}\rangle:=
    (b_1^\dagger)^{\widetilde{x}_1}
    \cdots
    (b_n^\dagger)^{\widetilde{x}_n}
    |\mathrm{vac}\rangle,
\]
which satisfies $ N|\psi_{{\rm alg},q}\rangle=q |\psi_{{\rm alg},q}\rangle$. 
It remains to prove that the energy comparison in
\textit{Step II} always finds a replacement that
does not decrease the energy. Fix the two non-integer eigenvalues
$\nu_\ell,\nu_m$ selected in \textit{Step I} and keep all
other eigenvalues unchanged.
Writing $\nu_\ell=t$ and $\nu_m=s-t$,
we show that the Gaussian energy of the modified intermediate state is a convex
function of $t$. To this end, using
$R=U\operatorname{diag}(\nu)U^\dagger$, we write
\[
    R_{jj}=\sum_{p=1}^n\nu_p|U_{jp}|^2,\quad
    R_{kk}=\sum_{p=1}^n\nu_p|U_{kp}|^2,\quad
    R_{jk}=\sum_{p=1}^n\nu_pU_{jp}\overline{U_{kp}}.
\]
Furthermore, from the expressions above and Proposition 
\ref{prop:wick-occupations}, we obtain 
\[
\begin{aligned}
    \langle n_jn_k\rangle_{\rho_R}=R_{jj}R_{kk}-|R_{jk}|^2
    &= \sum_{p,r=1}^n\nu_p\nu_r
\Bigl(
|U_{jp}|^2|U_{kr}|^2
-U_{jp}\overline{U_{kp}}
\overline{U_{jr}}U_{kr}
\Bigr)\\
&=\sum_{1\le p<r\le n}\nu_p\nu_r
       \Bigl(|U_{jp}U_{kr}|^2+|U_{jr}U_{kp}|^2
       -2\operatorname{Re}\!\left(
          U_{jp}U_{kr}\overline{U_{jr}U_{kp}}\right)\Bigr)\\
          &=\sum_{1\le p<r\le n}\nu_p\nu_r
       \left|U_{jp}U_{kr}-U_{jr}U_{kp}\right|^2.
\end{aligned}
\]
By~\eqref{eq:gaussian-energy}, the joint occupations
enter the Gaussian energy with coefficients
$-w_{jk}\leq 0$. Summing the terms involving
the product $\nu_\ell\nu_m$ over all edges gives
\[
    -C\nu_\ell\nu_m,\;\;\textup{where}\;\;
    C:=\sum_{\{j,k\}\in E}w_{jk}
    |U_{j\ell}U_{km}-U_{jm}U_{k\ell}|^2
    \geq 0.
\]
All other terms in the Gaussian energy are
either independent of $\nu_\ell,\nu_m$ or
linear in these eigenvalues, including the
hopping terms. 
Consequently, with all other eigenvalues fixed,
the Gaussian energy is a quadratic function
of $t$ of the form $ A+Bt-Ct(s-t)$, for some constants $A,B\in \R$.
Since $C\geq 0$, this function is convex. The constraints $0\leq\nu_\ell,\nu_m\leq 1$
imply that $t\in[\max\{0,s-1\},\min\{1,s\}]$. 
Notice that the two proposed replacements in \textit{Step I} correspond precisely
to the endpoints of this interval. 
By convexity, at least one endpoint yields an
energy no smaller than that of the current
Gaussian state. Hence, \textit{Step II} always
selects a replacement that does not decrease
the Gaussian energy.  Since each update preserves or increases the Gaussian
energy, successive iterations yield a nondecreasing
sequence of energies, starting from
$\langle H\rangle_{\rho_R}$ and ending at
$\langle H\rangle_{\psi_{{\rm alg},q}}$, hence the proof.
\end{proof}

\begin{corollary}
\label{cor:exact-particle-bipartite}
For $H$ in~\eqref{eq:H-particle-number}, suppose that the
interaction graph $(V,E)$ is bipartite and let $q\in \mathbb{N}$ satisfy $0\le q\le\left\lfloor \frac n2\right\rfloor$.  Consider the following algorithm:
\begin{enumerate}
    \item Construct the Gaussian state $\rho_{\mathrm{alg},\langle q\rangle}$
    following equation~\eqref{eq:state-selection-mean-q}. 
    \item Apply the procedure of
    Lemma~\ref{lem:fixed-particle-rounding} to its one-particle
    matrix, and return the resulting Slater determinant
    $|\psi_{\mathrm{alg},q}\rangle$.
\end{enumerate}
Then, the output $|\psi_{{\rm alg},q}\rangle$ has
exactly $q$ particles and satisfies
\[
    \langle H\rangle_{\psi_{\mathrm{alg},q}}
    \ge\frac12\operatorname{SDP}_{2,\langle q\rangle}(H)
    \ge\frac12\operatorname{OPT}_{\langle q\rangle}(H)
    \ge\frac12\operatorname{OPT}_q(H).
\]
\end{corollary}

\begin{proof}
The state $\rho_{\mathrm{alg},\langle q\rangle}$ constructed in
Theorem~\ref{thm:particle-number-bipartite} is number-conserving
Gaussian and satisfies $\langle N\rangle_{\rho_{\mathrm{alg},\langle q\rangle}}=q$.
Its one-particle matrix therefore has trace $q$
by~\eqref{eq:particle-trace}, so
Lemma~\ref{lem:fixed-particle-rounding} applies. Combining this with the guarantee of
Theorem~\ref{thm:particle-number-bipartite}, we obtain 
\[
    \langle H\rangle_{\psi_{\mathrm{alg},q}}
    \ge\langle H\rangle_{\rho_{\mathrm{alg},\langle q\rangle}}
    \ge\frac12\operatorname{SDP}_{2,\langle q\rangle}(H).
\]
Finally, the remaining inequalities follow from the fact that the SDP upper-bounds the mean particle-number optimum and that every state with exactly $q$
particles also has expected particle number $q$.
\end{proof}
Let us also mention that the postprocessing procedure of Lemma~\ref{lem:fixed-particle-rounding} also
recovers Lieb's variational principle for
Hartree--Fock theory~\cite{Lieb1981,Bach2022}
in the setting of our Hamiltonians:
\begin{equation}
\label{eq:Gaussian-Slater-equality}
\max_{\substack{
0\preceq R\preceq \mathbb{I},\\
\operatorname{Tr}(R)=q
}}
\langle H\rangle_{\rho_R}
=
\max_{\substack{
R^2=R,\\
\operatorname{Tr}(R)=q
}}
\langle H\rangle_{\rho_R},
\quad q\in\{0,\ldots,n\}.
\end{equation}
In other words, allowing a
number-conserving Gaussian state to have a prescribed mean particle number does not improve the optimal energy
over Slater determinants with exactly that many particles. Unlike the approximation guarantee in Corollary \ref{cor:exact-particle-bipartite}, this
equality does not require a bipartite interaction graph
and holds for every integer $q\in\{0,\ldots,n\}$.
%%%%%%%%%%%%%%%%%%%%%%%%%%%%%%%%%%%%%%%%%%%%%%%%%%%%
\section{Conclusion and Future Directions}
\label{sec:discussion}

In this work, we developed approximation algorithms for
classically interacting fermionic Hamiltonians by rounding level-two
fermionic moment relaxations to Gaussian states. Our approach combines
two complementary rounding procedures: one preserves quadratic
expectations, while the other exploits occupation correlations.

Using constraints involving only pairs of modes, we analyzed the
rounding procedures edge by edge and obtained improved approximation
guarantees for positive semidefinite classically interacting
fermionic Hamiltonians, as well as for Fermionic Max Cut. For traceless Hamiltonians, we provided an additive bound on the energy loss incurred by
Gaussian rounding and established an improved constant-factor approximation
on bipartite interaction graphs. We further extended this guarantee
to $q$-particle-constrained problems, including an exact prescribed
particle number, thereby addressing an open question raised
in previous work.

Since our approximation analysis considers one edge at a time,
an interesting question is whether stronger guarantees could
be obtained by exploiting consistency between overlapping
edges or by using higher-order moment relaxations.
As the latter become increasingly expensive, it would
also be worth investigating whether sparse relaxations
that exploit the structure of the interaction and
hopping graphs can provide constant-factor guarantees.

Another question is how the approximation guarantees
depend on the relative strength of the quadratic and
quartic terms, particularly in the weakly interacting
regime discussed in~\cite{STH}, where Gaussian states
approach optimality.
Finally, studying the integrality gaps of these
fermionic moment relaxations and comparing them
with the best energies achievable by Gaussian states
could help distinguish the limitations of the
relaxations, the restriction to Gaussian states,
and the rounding procedures developed here.

%%%%%%%%%%%%%%%%%%%%%%%%%%%%%%%%%%%%%%%%%%%%%%%%%%%%%%%%%%%%%%%%%%%%%%%%%%%%%%%
\section*{Acknowledgments}
This research was supported by the Australian Government through the Australian Research Council's Discovery Projects funding scheme (project DP250104201). H.F. acknowledges funding by UK Research and Innovation (UKRI) under the UK government’s Horizon Europe funding guarantee EP/X032051/1.

%%%%%%%%%%%%%%%%%%%%%%%%%%%%%%%%%%%%%%%%%%%%%%%%%%%%%%%%%
\paragraph{AI Statement.} LLMs assisted with exploring and checking proof
strategies and technical details. The structure and exposition of the manuscript are a result of the authors’ work. The authors are fully responsible for the mathematical claims and the final manuscript. 

\bibliographystyle{alpha}
\bibliography{References}

\end{document}